\documentclass[a4paper,pre,twocolumn,showkeys,superscriptaddress]{revtex4}
\usepackage[english]{babel}
\usepackage[utf8]{inputenc}
\usepackage{graphicx,epsfig,color,placeins}
\usepackage{latexsym,hyperref}
\usepackage{amsfonts}
\usepackage{amsmath}
\usepackage{textcomp}
\def\<{\langle}
\def\>{\rangle}
\def\{{\lbrace}
\def\}{\rbrace}
\def\({\left(}
\def\){\right)}
\def\beq{\begin{equation}}
\def\eeq{\end{equation}}

\begin{document}

\title{Kardar-Parisi-Zhang Universality in First-Passage Percolation:
  the Role of Geodesic Degeneracy}%

\author{Pedro C\'ordoba-Torres}
\affiliation{Depto. F\'{\i}sica Matem\'atica y de Fluidos, UNED, Spain}

\author{Silvia N. Santalla}
\affiliation{Depto. F\'{\i}sica y Grupo Interdisciplinar de Sistemas
  Complejos (GISC), Universidad Carlos III de Madrid, Spain}

\author{Rodolfo Cuerno}
\affiliation{Depto. Matem\'aticas y Grupo Interdisciplinar de Sistemas
  Complejos (GISC), Universidad Carlos III de Madrid, Spain}

\author{Javier Rodr\'{\i}guez-Laguna}
\affiliation{Depto. F\'{\i}sica Fundamental, UNED, Spain}

\begin{abstract}
We have characterized the scaling behavior of the first-passage
percolation (FPP) model on two types of discrete networks, the regular
square lattice and the disordered Delaunay lattice, thereby addressing
the effect of the underlying topology. Several distribution functions
for the link-times were considered. The asymptotic behavior
of the fluctuations for both the minimal arrival time and the lateral
deviation of the geodesic path are in perfect agreement with the
Kardar-Parisi-Zhang (KPZ) universality class regardless of the type of
the link-time distribution and of the lattice topology. Pre-asymptotic
behavior, on the other hand, is found to depend on the uniqueness of
geodesics in absence of disorder in the local crossing times, a
topological property of lattice directions that we term \emph{geodesic
  degeneracy}. This property has important consequences on the model,
as for example the well-known anisotropic growth in regular
lattices. In this work we provide a framework to understand its effect
as well as to characterize its extent.
\end{abstract}


\date{February 13, 2018}

\maketitle
%%%%%%%%%%%%%%%%%%%%%%%%%%%%%%%%%%%%%%%%%%%%%%%%%%%%%%%%%%%%%%%%%%%%%%%%%%%%%%%

\section{Introduction}

Stochastic geometry presents a wealth of results, both from the
mathematical and the physical standpoints
\cite{Adler,Itzykson_Drouffe,Booss.book}. The physics of polymers,
membranes and fluctuating interfaces is described by random geometry
\cite{Nelson,Boal}, as is quantum gravity in two dimensions
\cite{Ambjorn_97}. Recently, it was shown that random two-dimensional
Riemannian manifolds endowed with a random metric field which is a
short-range perturbation of the plane metric show universal fractal
properties \cite{Santalla_15}. Straight lines and circumferences,
i.e. geodesics and balls, become irregular and their roughness follows
scaling laws within the \emph{Kardar-Parisi-Zhang} (KPZ)
\emph{universality class} \cite{Kardar_PRL86} which describes
interfacial random growth \cite{Barabasi,Krug_97,Krug_10,Halpin_15}.
Concretely, the width of a ball with radius $R$ can be shown to scale
as $W\sim R^\beta$, where $\beta=1/3$ is the growth exponent, and the
lateral deviation of a geodesic between two points whose Euclidean
distance is $L$ scales as $L^{1/z}$, where $z=3/2$ is the dynamical
exponent. Moreover, the radial fluctuations at any point of a ball
were shown to follow the \emph{Tracy-Widom distribution} associated
with the \emph{Gaussian unitary ensemble} (TW-GUE)
\cite{Praehofer_02,Takeuchi_11,Corwin_13}. The same calculation was
performed using other base manifolds, instead of the Euclidean plane
\cite{Santalla_17}. For example, for a cylinder, the KPZ class is
again found, but this time the radial fluctuations follow the
\emph{Gaussian orthogonal ensemble} (TW-GOE).

In this work we consider the {\em first-passage percolation} (FPP)
model \cite{Hammersley_65,Howard_04,Auffinger_17}, the classic
discrete model of fluid flow through a random medium whose continuum
counterpart is the random metric model of Refs.
\cite{Santalla_15,Santalla_17}. The discrete representation presents a
whole set of new phenomena which ask for a thorough understanding. The
FPP model consists of a network in which each nearest-neighbor link is
endowed with certain random crossing time. Given two nodes of the
lattice, we can compute the \emph{minimal arrival time} required to
travel between them, and then the associated \emph{minimal-time path},
i.e. the {\em geodesic}. If the link-times are allowed to
fluctuate, then the arrival time will fluctuate too. Indeed, a small
change in the link times may cause a large change in the minimal-time
path. Thus, the statistics of minimal paths and minimal arrival times
are strongly associated.

The FPP problem has been thoroughly studied on Bernoulli systems, when
the link-times can only be zero or one
\cite{Smythe_78,Ritzenberg_84,Kerstein_86,Yao_13,Yao_18}, and on
higher dimensions \cite{Fill_93}, including random graphs and small
worlds \cite{Bhamidi_10}. Indeed, the so called KPZ relation between
the scaling exponents, $z(1+\beta)=2$, has been proved for FPP balls
in any dimension \cite{Chatterjee_13} provided that the exponents are
suitably defined \cite{Auffinger_17}. Moreover, the FPP problem bears
relation to the study of directed polymers in random media (DPRM)
\cite{Kardar_87,Krug_91,Halpin_95}. Results about FPP have found
applications in areas as distant as magnetism \cite{Abraham_95},
wireless communications \cite{Beyme_14}, ecological competition
\cite{Kordzhakia_05} or sequence alignment in molecular biology
\cite{Bundschuh_00}.

\begin{figure*}
  \includegraphics[width=7cm]{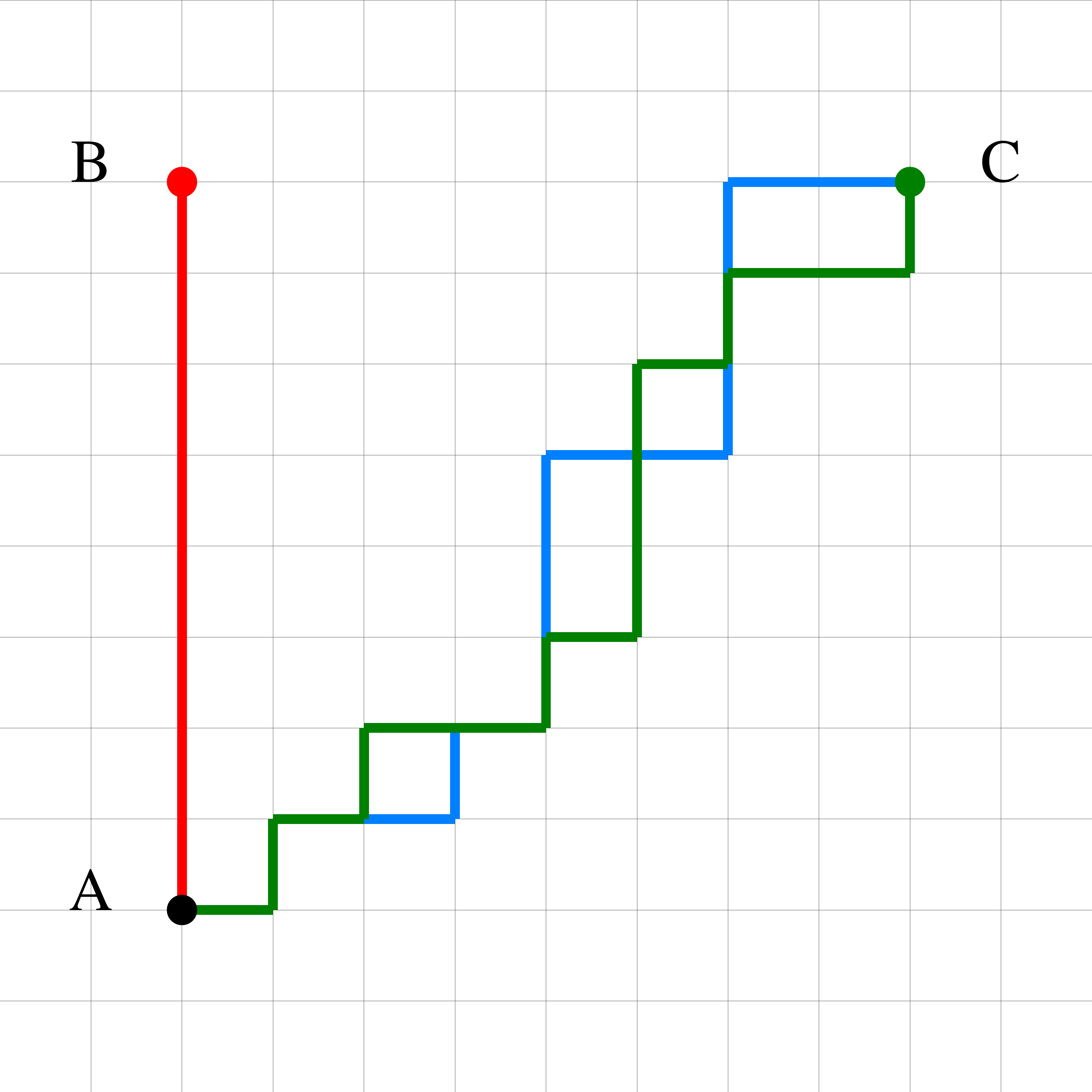}
  \hspace{1cm}
  \includegraphics[width=7cm]{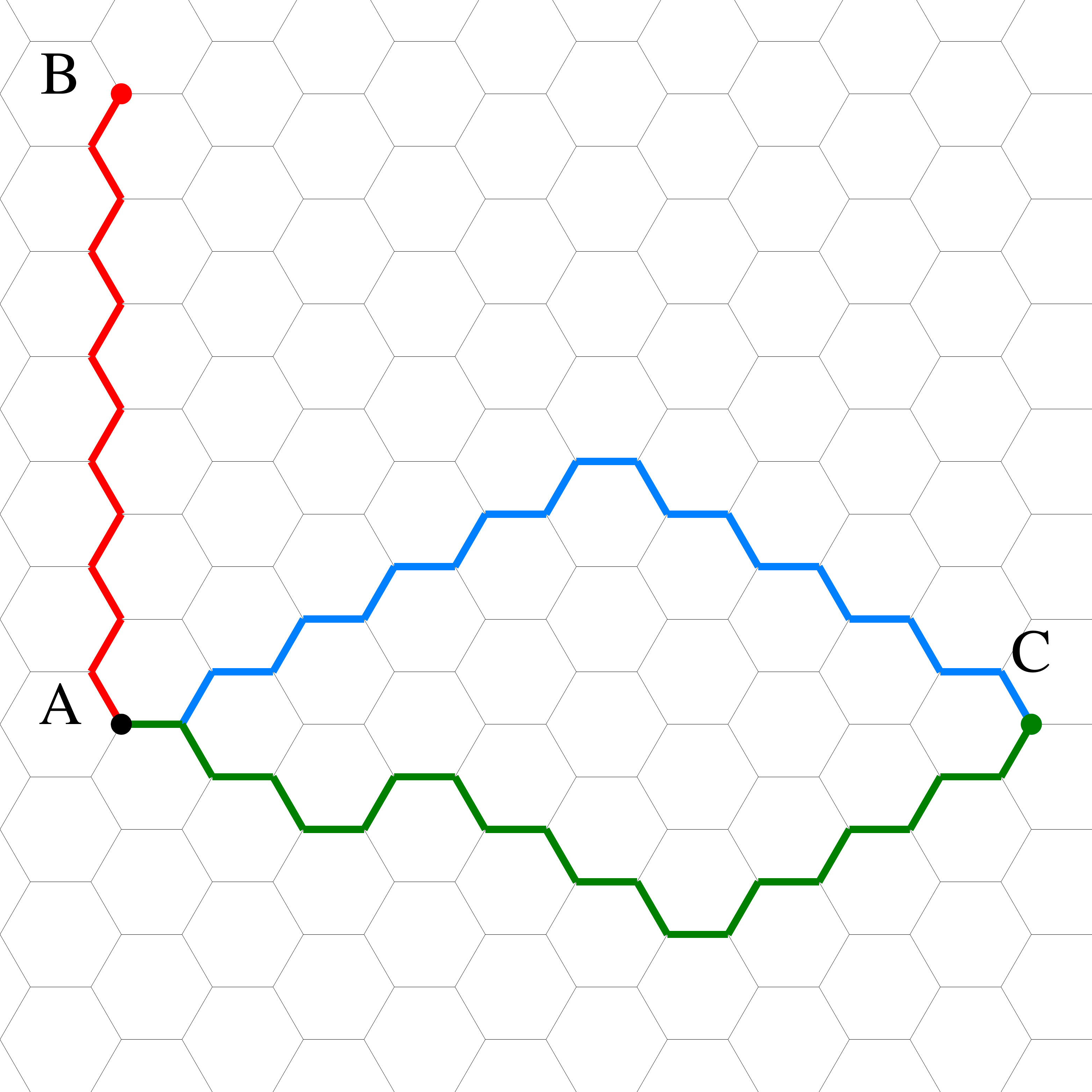}
  \caption{Illustrating the concept of geodesic degeneracy on the
    square (left) and hexagonal (right) lattices. We consider the
    homogeneous case in which link-times are equal for all
    links. The geodesic (minimal-time path) between points A and B is
    unique in both cases, while it is strongly degenerate when points
    A and C are considered.}
  \label{fig:illust_degeneracy}
\end{figure*}

We focus on an interesting property of lattice directions, not yet
addressed. In the case where all link-times are equal (the
so-called \emph{homogeneous} or \emph{clean case}), the geodesic
between any pair of lattice nodes may be unique or otherwise it may be
be {\em degenerate}, i.e. there is a number of different minimal-time
paths connecting the two nodes all having equal (minimal) arrival
times. This property, to which we will refer to as \emph{geodesic
  degeneracy}, depends on the lattice direction, so that anisotropic
behavior is expected in regular lattices. This is illustrated in
Fig. \ref{fig:illust_degeneracy} for the square and hexagonal
lattices. In both cases points A and B are joined by a single geodesic
in the homogeneous limit. Thus, that lattice direction has no geodesic
degeneracy. On the contrary, there is a large number of minimal paths
between A and C (only two examples have been highlighted), hence that
specific lattice direction is strongly degenerate. As we will show in
this paper, this property is very relevant to understand how the
geodesics and the times of arrival fluctuate when the link-times are allowed to vary, specially in the pre-asymptotic
regime. 

We have performed a thorough numerical analysis of the arrival time
statistics and geodesic geometry on two types of discrete planar
lattices, a regular one given by the square lattice and a disordered
Delaunay lattice formed from random points on the Euclidean
plane. Delaunay lattices are triangulations which fulfill a very
stringent constraint: none of the triangle circumcircles may contain
any other lattice point. In addition, for a comprehensive
characterization of the model we have considered several distribution
functions for the link-times. Our results provide strong
evidence that FPP in planar lattices falls asymptotically into the KPZ
universality class. Moreover, beyond the asymptotic behavior, we were
able to characterize the pre-asymptotic regime and the crossover time
in both types of lattice, which may be of practical importance for
specific applications of the FPP model. Note that finite simulations
can well be dominated by the preasymptotic behavior, as it is common
in the context of scale-invariant processes.

This article is organized as follows. Section \ref{sec:model}
describes the FPP model on the square lattice, as well as the
definitions employed in the article. In section \ref{sec:time_fluct}
we characterize the fluctuation of the times of arrival to points
along the axis and the diagonal of the square lattice, showing that
the origin of their difference stems from their geodesic
degeneracy. Then, in section \ref{sec:geodesics} we consider actual
geodesics, specifically their lateral deviation, and show how the same
concept allows for a complete characterization. The shapes of the
growing balls for long times is the focus of
Sec. \ref{sec:limitshape}, where we address the anisotropic growth
caused by the lattice anisotropy in the geodesic degeneracy. In order
to distinguish the idiosincracies of the square lattice from more
general features of the model, we have studied random triangulations
of the plane by tracing stochastic Delaunay lattices. The results, as
shown in Sec. \ref{sec:delaunay}, are consistent with those found for
the square lattice once we consider the corresponding geodesic
degeneracy. Finally, Sec. \ref{sec:conclusions} summarizes our
conclusions and discusses interesting lines of future work.

%%%%%%%%%%%%%%%%%%%%%%%%%%%%%%%%%%%%%%%%%%%%%%%%%%%%%%%%%%%%%%%%%%%%%%%%%%%%

\section{Model}
\label{sec:model}

\subsection{First-passage percolation: arrival times and geodesics}

Let us consider an undirected graph ${\cal L}$ with $N$ nodes and a
given center node $\textbf{x}_0$. A link-time
$t(\textbf{x}_i,\textbf{x}_j)$ is associated with each link between
nearest-neighbor nodes $\textbf{x}_i$ and $\textbf{x}_j$. Now, we find
the {\em minimal arrival time} $T(\textbf{x})$ (also referred to as
\emph{passage time}) from the center node $\textbf{x}_0$ to all other
nodes on the lattice $\textbf{x}\in {\cal L}$:
\beq
T(\textbf{x})=\min_{m,\{\textbf{x}_1,\cdots,\textbf{x}_{m-1}\}}
\sum_{i=1}^{m} t(\textbf{x}_{i-1},\textbf{x}_{i}),
\label{eq:minimalarrivaltime}
\eeq
where we assume that $\textbf{x}_m=\textbf{x}$ and
$t(\textbf{x},\textbf{y})=\infty$ if $\textbf{x}$ and $\textbf{y}$ are
not nearest-neighbors. Notice that the length of the path, $m$, is
also minimized. This minimal arrival time can be obtained
using e.g. Dijkstra's algorithm \cite{Cormen}, which works in $O(N^2)$
time for an arbitrary graph. Besides the minimal arrival time,
Dijkstra's algorithm also returns the {\em parent} of each node,
$P(\textbf{x})$, which is the node from which $\textbf{x}$ is reached
when the minimal-time path is followed. Thus, by applying iteratively
the parent application we eventually must reach the center node:
\beq
\forall \textbf{x} \in {\cal L},\quad \exists \, !n
\quad |\quad P^{(n)}(\textbf{x})=\textbf{x}_0,
\label{eq:rootnode}
\eeq
and we call $n(\textbf{x})$ the {\em degree} of node
$\textbf{x}$. Note that $n(\textbf{x})$ is the value of $m$ resulting
from the minimization in Eq. (\ref{eq:minimalarrivaltime}). In this way
the {\em geodesic} associated to that node (also known as \emph{optimal path}) is the orbit obtained from the successive application of $P$:
\beq
G(\textbf{x})=\{ P^{(k)}(\textbf{x}):\quad k=1,\ldots,n(\textbf{x})  \},
\label{eq:geodesic}
\eeq
It must be stressed that we have assumed that the value of $n$ is
unique for each lattice node, which means that the geodesic path
between lattice points is unique too. It seems to be a reasonable
assumption when the distribution of the link times is
\emph{continuous}.

For regular lattices with constant spacing (as those illustrated in Fig. \ref{fig:illust_degeneracy}), the length of the geodesic
path in lattice units, denoted by $l$, will be given by:
\beq
l(G(\textbf{x}))=n(\textbf{x}).
\label{eq:geodesic_lenght}
\eeq
Finally, we can also define an open {\em ball} as the set of nodes which can be
reached in a time smaller than a certain value $t$:
\beq
B(t)=\{\textbf{x}\in {\cal L}:\quad T(\textbf{x})<t\}.
\label{eq:balls}
\eeq

\subsection{Geometric setup and link times}

The first set of results reported here were obtained in a square
lattice of lateral size $2L+1$ and with $\textbf{x}_0$ at its
geometrical center. The choice of this simple geometry obeys two
purposes. From the technical point of view, it allows a large number
of simulations using large lattice sizes in order to accurately
characterize the asymptotic scaling of the fluctuations. Unless
otherwise stated, displayed results were obtained for $L=1000$ and
from $2.5\cdot 10^4$ simulations, which turned into $10^5$ points due
to the $\pi/2$ rotational symmetry of the lattice.

From the fundamental point of view, the structure of the square
lattice lends itself to a careful study of the effect of
\emph{geodesic degeneracy}, as described in the Introduction. Indeed,
let us consider two nodes separated by vector $(x,y)$. In the case
where all link times are equal (the so-called \emph{homogeneous
  case}), the length of the optimal (geodesic) path between them will
be $l=|x|+|y|$. However, this optimal path is typically
\emph{degenerate}, and the number of different geodesics connecting
the two nodes is given by
\beq
N_{deg}(x,y)={(|x|+|y|)!\over |x|!\;|y|!},
\label{eq:number_geodesics_homoheneous}
\eeq
which we can call the \emph{degree of degeneracy} associated to the
direction $(x,y)$. For a constant geodesic length, say $l=2\ell$, the
highest degree of degeneracy is obtained when the sites are on a
lattice diagonal ($|x|=|y|=\ell$) and it is given by $N_{deg,
  \text{max}}=(2\ell)!/ (\ell!)^2 \approx 2^{2\ell}$. On the
other side, the lowest degeneracy corresponds to points on an axis
($\pm 2\ell,0)$ or ($0,\pm 2\ell$) resulting in $N_{deg,
  \text{min}}=1$, which means that the geodesic path is \emph{unique}
and given by the Euclidean straight line connecting the two sites. For
intermediate lattice directions degeneracy increases with their angle
with respect to the axis. The amount of geodesic degeneracy is an
intrinsic property of the lattice, and the maximal exponential growth
rate is given by the maximal eigenvalue of the associated adjacency
matrix.

Most of our work will compare distances between nodes and times of
arrival, so we define $d(\textbf{x}_i,\textbf{x}_j)$ as the
standard Euclidean distance between nodes $\textbf{x}_i$ and
$\textbf{x}_j$. For the square lattice we will assume that path lengths (denoted by
$l$) and distances between lattices sites (given by $d$) will be given in units
of the lattice spacing. It is worth noticing that times of arrival can
be regarded as distances in a different metric, as it is done in
\cite{Santalla_15}. Disordered lattices will be addressed in section
\ref{sec:delaunay} and their construction will be discussed there.

Once the lattice is set, we provide each nearest-neighbor link with a
crossing time $t$, which is randomly and independently chosen from a
distribution function $f(t)$ with mean $\tau$ and variance $s^2$
(standard deviation $s$). We have worked with different distribution
functions such as uniform, log-normal, Weibull, and Pareto. Let us
describe briefly our choice of parameters. Uniform distributions on an
interval $[t_{\text{min}},t_{\text{max}}]$ will be denoted by
$\text{U}(t_{\text{min}},t_{\text{max}})$ with the following relations
holding:
\beq
t_{\text{min}}=\tau-\sqrt{3}s, \qquad t_{\text{max}}=\tau+\sqrt{3}s.
\label{eq:tminmax}
\eeq
We can define the amplitude of the fluctuations as $\delta_t\equiv t_{max}-t_{min}$ and we obtain
\beq
\delta_t=2\sqrt{3}s.
\label{eq:amplitude_fluct}
\eeq
Log-normal distributions will be denoted as LogN$(\mu,\sigma^2)$, where
the distribution pdf is given by:
\beq
f(t)={1\over t\sigma\sqrt{2\pi}} \exp\(-(\ln t-\mu)^2\over 2\sigma^2\).
\label{eq:logn}
\eeq
The Weibull distribution is denoted here by Wei$(\lambda,k)$ with
pdf (only defined for positive $x$):
\beq
f(t)={k\over\lambda} \({t\over\lambda}\)^{k-1}\exp\(-(t/\lambda)^k\).
\label{eq:weibull}
\eeq
Finally, the Pareto distribution will be termed as Par$(t_m,\alpha)$, defined for
$t>t_m$, with $t_m$ and $\alpha>0$, and with the following pdf:
\beq
f(t)={\alpha t_m^\alpha\over t^{\alpha+1}}.
\label{eq:pareto}
\eeq

%%%%%%%%%%%%%%%%%%%%%%%%%%%%%%%%%%%%%%%%%%%%%%%%%%%%%%%%%%%%%%%%%%%%%%%%%%%%

\section{Fluctuations of the Minimal Time of Arrival}
\label{sec:time_fluct}

We begin our analysis by addressing the fluctuations of the minimal
time of arrival to the nodes of the square lattice as a function of
the distance to the center node $\textbf{x}_0$, which is the origin of
coordinates. Let us remark that fluctuations in the minimal arrival
time correspond to the {\em roughness} of the balls
\cite{Chatterjee_13,Santalla_15}, and will be characterized by the
same scaling exponent, $\beta$, as long as it exists. Thus, within the
KPZ class we expect the variance of the minimal time of arrival
$\sigma^2_T \sim d^{2\beta}$, where $\beta=\frac{1}{3}$.

As discussed above, the structure of the square lattice suggests
focusing the analysis on two lattice directions, the \emph{axis} and
the \emph{diagonal}, as they stand for the limiting cases from which
intermediate behavior should be readily deduced.

\subsection{Scaling on the axis}

\begin{figure*}[ht]
  \includegraphics[width=16cm]{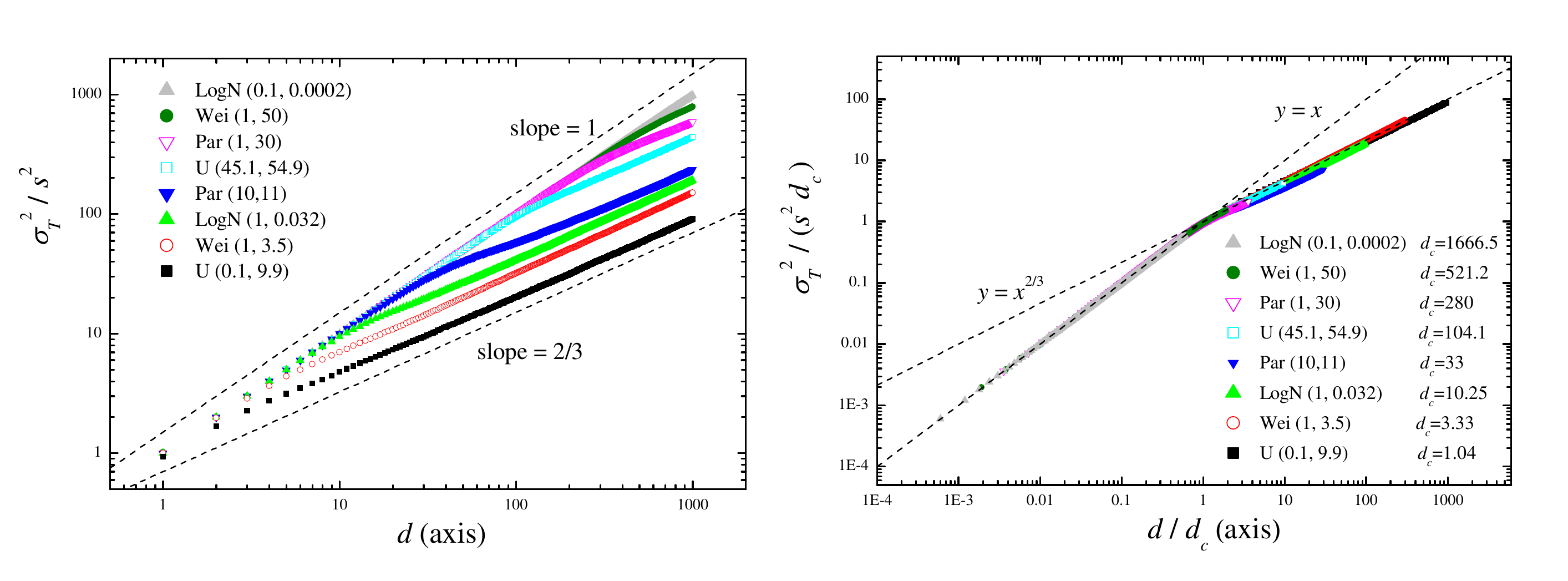}
  \caption{Fluctuations of the minimal time of arrival to points on
    the axis of the square lattice as a function of their Euclidean
    distance to the origin, for different link-time
    distributions. (Left) Fluctuation variance has been rescaled by
    the link-time variance $s^2$. Broken lines represent power-law
    behaviors with exponents 1 and 2/3, as indicated. (Right) Distance
    and variance have been rescaled by the corresponding crossover
    distance $d_c$, indicated for each link-time distribution in the
    legend. Broken lines represent the two branches of the piecewise
    scaling function given in Eq.  (\ref{eq:scaling_function_axis}).}
  \label{fig:fluct_axis}
\end{figure*}

We consider the times of arrival to points $\textbf{x}$ on the axis,
i.e. points with coordinates of the form $(\pm x,0)$ and $(0,\pm x)$,
with $x=1,\ldots,L$, and whose Euclidean distance to the origin is
$d(\textbf{x}_0,\textbf{x})=x$. We show in Fig. \ref{fig:fluct_axis}
(left) the variance of the minimal time of arrival, $\sigma_T^2$,
rescaled by the link-time variance $s^2$, as a function of the
distance to the origin for different distribution functions and some
representative parameters. Two different scaling regimes indicated by
the broken lines are clearly observed. For most cases there is an
initial regime of the form $\sigma_T^2 \sim d$, which is followed by
the asymptotic scaling $\sigma_T^2 \sim d^{2\beta}$ with
$\beta=\frac{1}{3}$, in agreement with the expected KPZ universality
class \cite{Santalla_15}. The pre-asymptotic regime can be arbitrarily
large and exceed the lattice limits, as in the upper curve
corresponding to LogN$(0.1,0.0002)$, or arbitrarily short so that it
can not be observed, e.g. lower curve, U$(0.1,9.9)$.

The reason for the pre-asymptotic linear regime $\sigma_T^2\sim d$ is
the following. As discussed above, the optimal path between two points
on the axis in the uniform case $s^2=0$ is unique. When $s^2$ is
positive but very small, still the geodesic will correspond to the
Euclidean line because the deviation from it entails additional steps
(at least two) which come at a cost in time proportional to $\tau$,
the mean link-time. For short distances this is enough to preclude
any deviation of the minimal-time path from the axis. In this regime
the average passage time between two points separated by a distance
$d$ is simply $\langle T(d)\rangle = d\tau$, and its variance comes
from the straightforward addition of the link-time variances
$\sigma_T^2(d) = ds^2$, thereby accounting for the pre-asymptotic
scaling displayed in Fig. \ref{fig:fluct_axis} (left).

The amplitude of the arrival-time fluctuations will grow with distance
$d$ until it becomes large enough to assume the cost in time of an
eventual deviation from the Euclidean geodesic. We will denote that
critical distance by $d_c$. For distances above $d_c$ the disorder
amplitude makes the underlying geometric constraint imposed by the
lattice irrelevant and hence allows the geodesics to explore freely the
space.

We can deduce an accurate expression for the crossover distance if we first assume Eq. (\ref{eq:amplitude_fluct}) for the amplitude of the minimal time fluctuations $\delta_T(d)$,
\beq
\delta_T(d)=2\sqrt{3\sigma_T^2(d)}=2s\sqrt{3d},
\label{eq:amplitude_fluctuations}
\eeq
and we note that the smallest deviation of the geodesic from the
axial line necessarily entails two additional steps which, on average,
represent a contribution of $2\tau$ to the passage time. After
equating the amplitude of the disorder at $d_c$ to that cost, $\delta_T(d_c)=2\tau$, we obtain
\beq
d_c = {\tau^2\over 3s^2} = \frac{1}{3} \frac{1}{(\text{CV})^2}.
\label{eq:def_dc}
\eeq
where CV is the {\em coefficient of variation} of the link-time
distribution, defined for every $f(t)$ as the ratio of the
standard deviation $s$ to the mean value $\tau$. This parameter is
frequently used in statistics as a standardized measure of the
dispersion of a distribution.

The expression given in Eq.(\ref{eq:def_dc}) agrees with our
qualitative description since $d_c$ grows with $\tau$ and decreases
with $s^2$. Moreover, it also agrees with the results displayed in
Fig. \ref{fig:fluct_axis} (left) as it predicts a value of $d_c=1.04$
for case U$(0.1,9.9)$, which thus precludes the observation of the
pre-asymptotic regime, and a value of $d_c=1666.5$ for case
LogN$(0.1,0.0002)$, which indicates that the lattice size ($L=1000$) is not
sufficiently large to reach the asymptotic KPZ scaling.
%From Eq. (\ref{eq:tminmax}) it can be readily shown that $d _c > 1$ for the
%uniform distribution.

The validity of Eq. (\ref{eq:def_dc}) is demonstrated in the right panel of
Fig. \ref{fig:fluct_axis}, where data have been rescaled by $d_c$ and
the curves collapse to a single universal function. That means that
fluctuations of the minimal time of arrival to nodes on the axis are
completely determined by the dispersion of $f(t)$, concretely by its
coefficient of variation, so that different distribution functions but
with the same CV will yield similar behaviors. We thus deduce the following scaling Ansatz:
\beq
\sigma^2_T(d) = 3^{-1} \tau^2 g\( {d\over 3^{-1}\tau^2s^{-2}} \),
\label{eq:scaling_axis}
\eeq
with the scaling function
\beq
g(x)\sim \begin{cases} x & \textrm{if } x\ll 1, \\
  x^{2\beta} & \textrm{if }x\gg 1. \end{cases}
\label{eq:scaling_function_axis}
\eeq

\begin{figure}[ht]
  \includegraphics[width=\columnwidth]{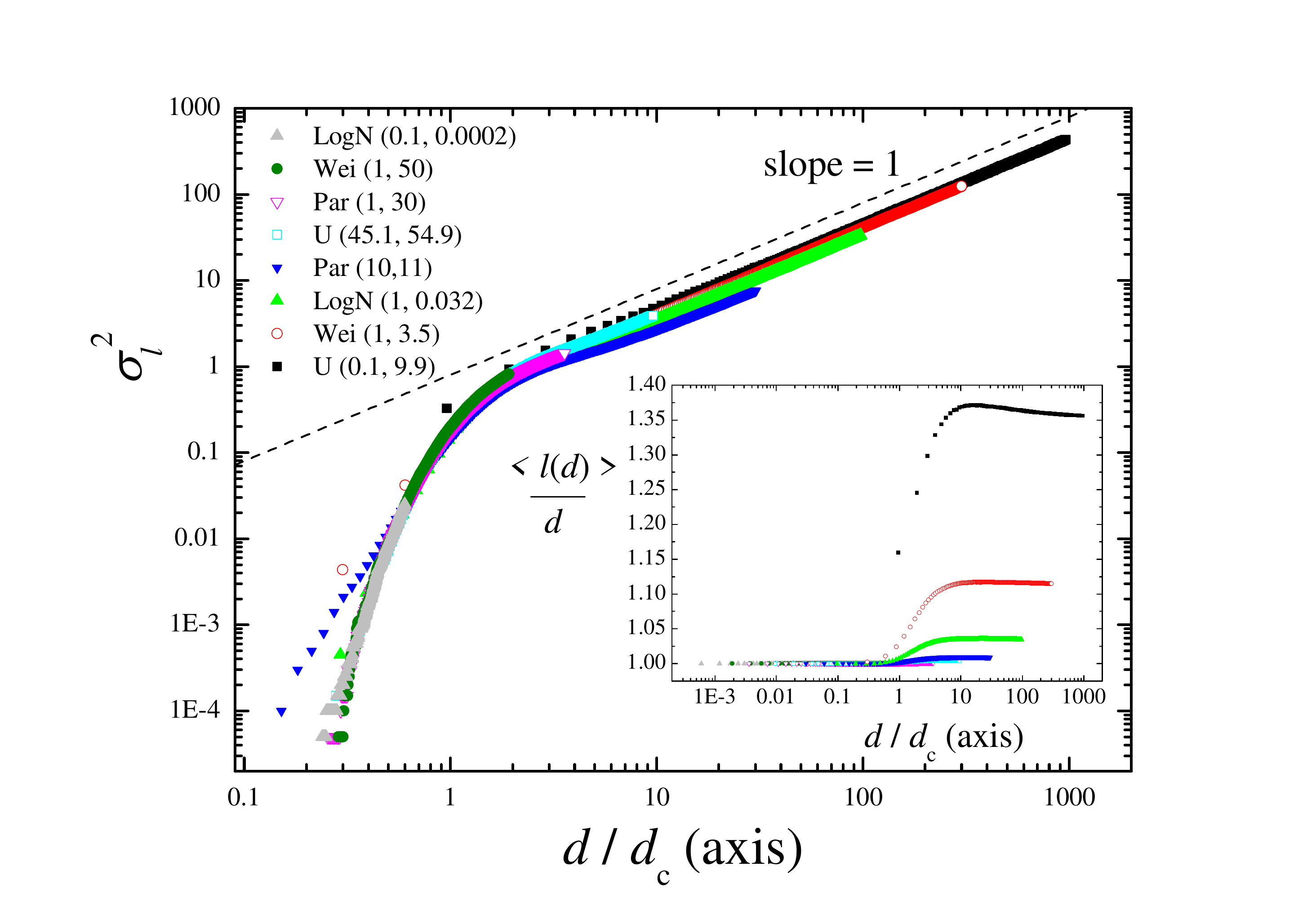}
  \caption{Variance of the geodesic length as a function of the scaled
    distance on the axis for the same set of results displayed in
    Fig. \ref{fig:fluct_axis}. The broken line indicates the linear
    behavior. (Inset) Corresponding average geodesic length rescaled
    by the Euclidean distance. }
  \label{fig:fluct_length_axis}
\end{figure}

To add more consistency to our reasoning we have displayed in
Fig. \ref{fig:fluct_length_axis} the variance of the length of the
geodesic path, $\sigma^2_l$, as a function of the rescaled
distance. An excellent collapse to the following scaling function is
again obtained:
\beq
\sigma^2_l(d) \sim q\(d\over d_c \),
\label{eq:scaling_length_axis}
\eeq
with
\beq
q(x)\sim \begin{cases} 0 & \textrm{if } x\ll 1 \\
  x & \textrm{if } x\gg 1 \end{cases}.
\label{eq:scaling_function_length_axis}
\eeq
In the pre-asymptotic regime ($d\ll d_c$) optimal paths follow the
Euclidean axis and $\sigma^2_l(d)= 0$. Above $d_c$, the variance of
the geodesic length increases with distance because the increase of
the fluctuations allows the geodesics to explore the space, now free
of geometrical constraints, in more complex ways. The distribution of
the minimal-path length fluctuations approaches the normal
distribution as $d$ increases so that the scaling $\sigma^2_l(d) \sim
d$ seems to result from the sum of uncorrelated random
variables. Details on the behavior of the average geodesic length
$\<l(d)\>$ (scaled by the Euclidean length $d$) are displayed in the
inset of Fig. \ref{fig:fluct_length_axis}. As expected, for $d\ll d_c$
we have $\<l(d)\>=d$ whereas for $d \gg d_c$ the ratio seems to
approach a constant value that increases with the CV of $f(t)$.

\subsection{Scaling on the diagonal}

\begin{figure}[ht]
  \includegraphics[width=\columnwidth]{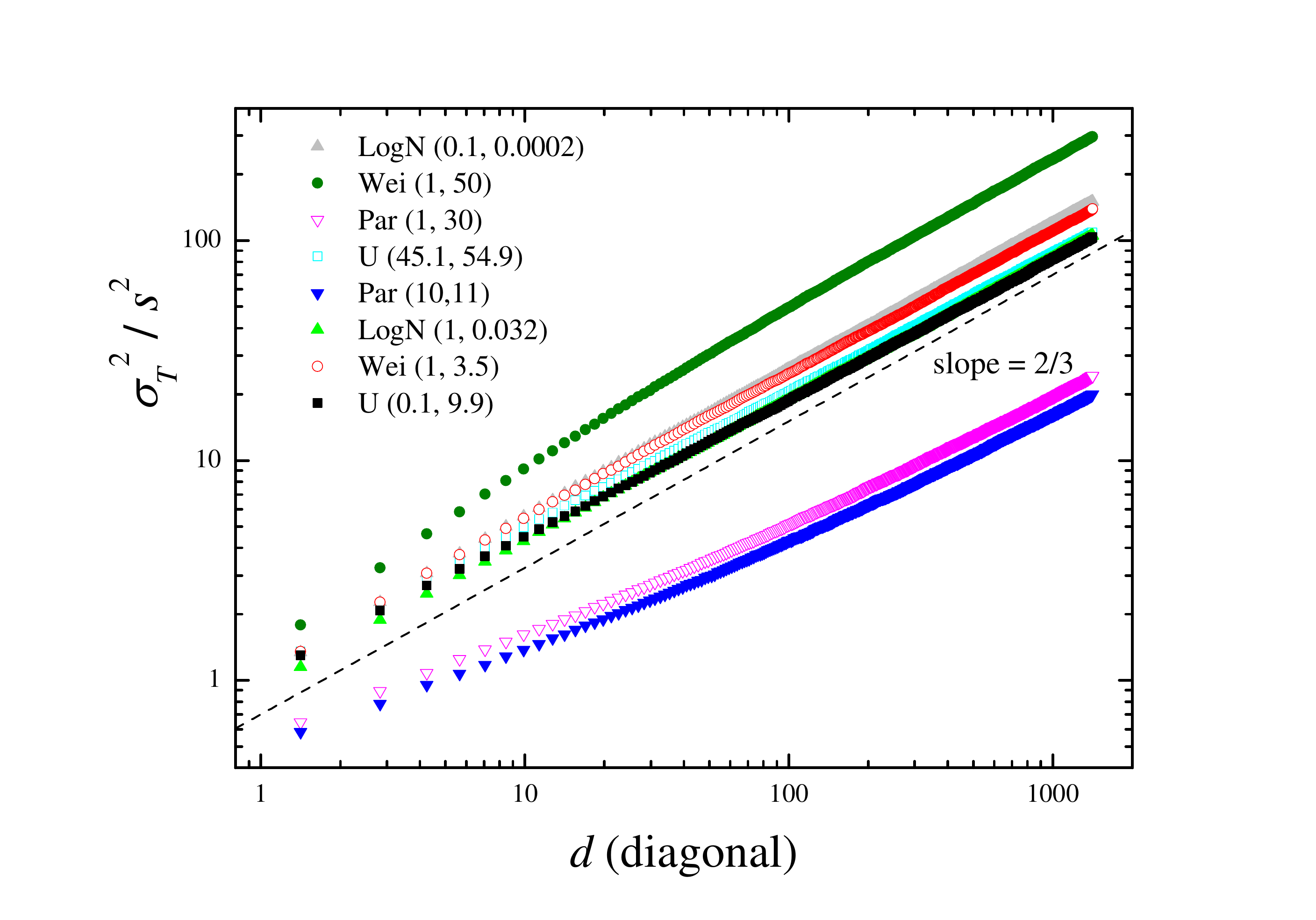}
  \caption{Scaled variance of the fluctuations of the minimal arrival
    time to points on the diagonal of the square lattice as a function
    of their Euclidean distance to the origin, for the same
    link-distributions considered in Fig. \ref{fig:fluct_axis}. The
    broken line stands for the KPZ scaling.}
  \label{fig:fluct_diag}
\end{figure}

Let us now consider the behavior along the lattice diagonals,
i.e. corresponding to points $\textbf{x}$ whose coordinates are of the
form $(\pm x, \pm x)$ with $x=1,\ldots,L$, and Euclidean distance to
the origin given by $d(\textbf{x}_0,\textbf{x})= \sqrt{2}x$. The
growth of the variance of the minimal-time fluctuations has been
displayed in Fig. \ref{fig:fluct_diag} for the same link-time
distributions considered in Fig. \ref{fig:fluct_axis}. Contrary to the
axis, the pre-asymptotic regime has nearly disappeared and
fluctuations start KPZ scaling (broken line) at very early
times. Moreover, the small remaining transient seems to depend on the
type of distribution function, being more marked for the Pareto
distribution.

This is a striking result since KPZ scaling is rapidly attained even
for distributions with a very small coefficient of variation (large
value of $d_c$), e.g. LogN$(0.1,0.0002)$, for which we had obtained a
trivial growth along the axis. The reason was introduced above being
the degeneracy of the geodesics in the homogeneous system. When
$s^2=0$ the minimal-time path between two points on the diagonal is
degenerate. As soon as the link-time distribution is introduced on the
lattice, the degeneracy is broken. However, for systems with low
dispersion (CV$\ll1$) the optimal path will be one of the geodesics of
the $s^2=0$ case, with fixed length $l(\textbf{x})=2x$. Since
degeneracy increases exponentially with the distance, at short
distances the number of degenerate optimal paths will be large enough
to allow the minimal arrival time to fluctuate without the geometrical
constraints found along the axis direction.

\begin{figure}[ht]
  \includegraphics[width=\columnwidth]{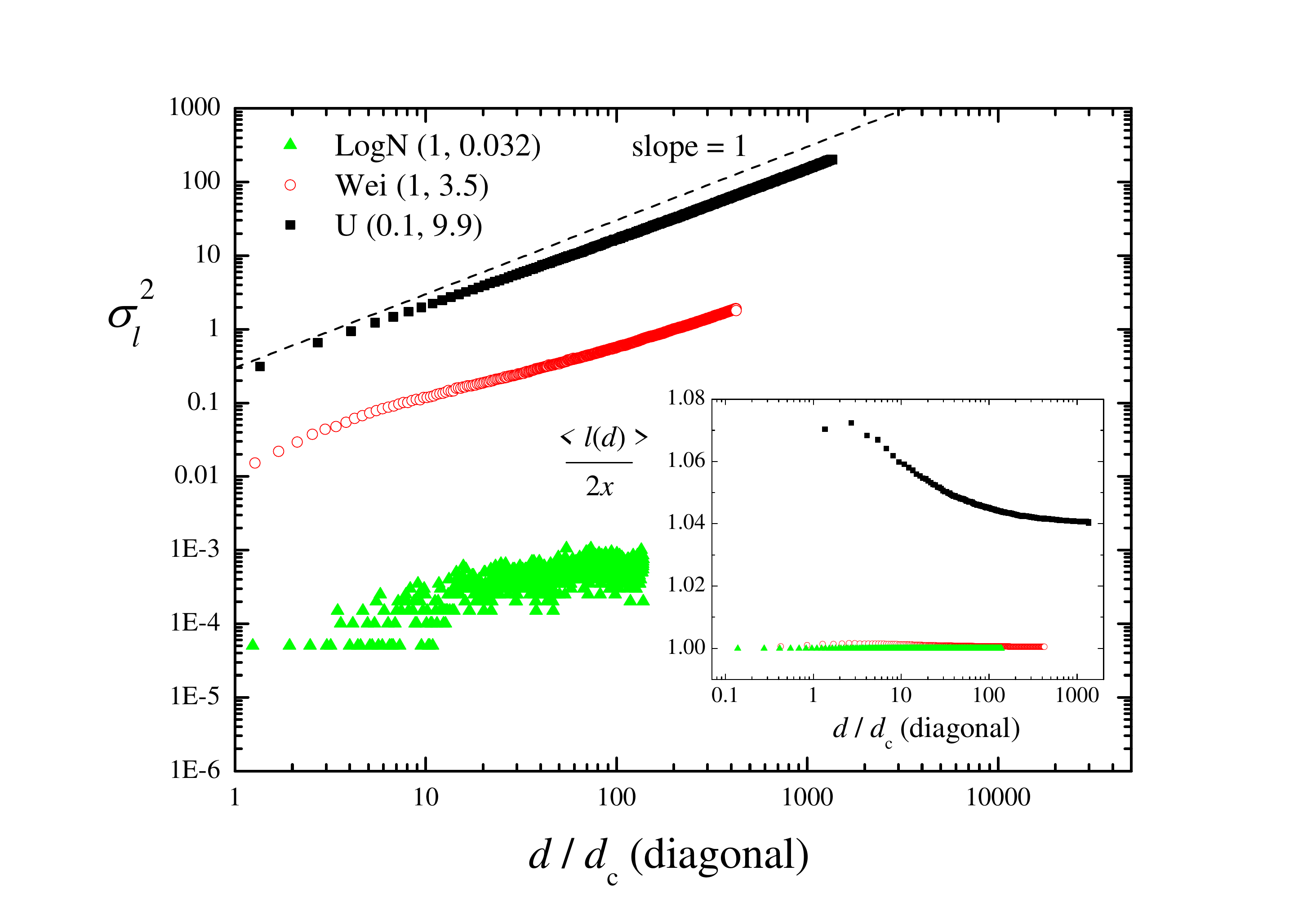}
  \caption{Variance of the geodesic length as a function of the scaled
    distance along the diagonal. The cases shown are the same as those
    displayed in Fig. \ref{fig:fluct_diag}, removing those which
    present zero fluctuations. The broken line indicates the
    linear behavior. (Inset) Corresponding average geodesic length
    scaled by the $l=2x$ value of case $s^2=0$.}
  \label{fig:fluct_length_diag}
\end{figure}

As expected, the effect of degeneracy is also noticeable in the
behavior of the geodesic length, whose variance has been displayed in
Fig. \ref{fig:fluct_length_diag}. Only those cases yielding non-zero
fluctuations (largest values of CV) have been plotted. For
distributions with low values of CV (say, $d_c>100$), the minimal-time
path was always one of the degenerate geodesics of the homogeneous
case $s^2=0$ so no length fluctuations were observed.

In the axis direction, KPZ scaling was directly related to the
deviation of the geodesics from the Euclidean path. Along the
diagonal, however, the KPZ behavior displayed in
Fig. \ref{fig:fluct_diag} has no relation to the fluctuations of the
geodesic length, which show no universal features. As discussed above,
the geodesic degeneracy prevents the optimal paths in the disordered
case from leaving the degenerate ensemble. Only when the amplitude of
the disorder is large enough to conceal the underlying lattice
structure, non-negligible fluctuations of the geodesic length are
observed. This happens when $d\gg d_c$ and becomes significant for
very large values of CV. Asymptotic behavior seems to follow the same
scaling as for the axis, $\sigma^2_l(d) \sim d$ (broken line in
Fig. \ref{fig:fluct_length_diag}). The average geodesic length (scaled
by the clean-case length $2x$) is displayed in the inset and shows a
similar behavior, although less pronounced, to the curves displayed in
the inset of Fig. \ref{fig:fluct_length_axis} for the axis.

\subsection{Full probability distribution}

\begin{figure} [ht]
  \includegraphics[width=9cm]{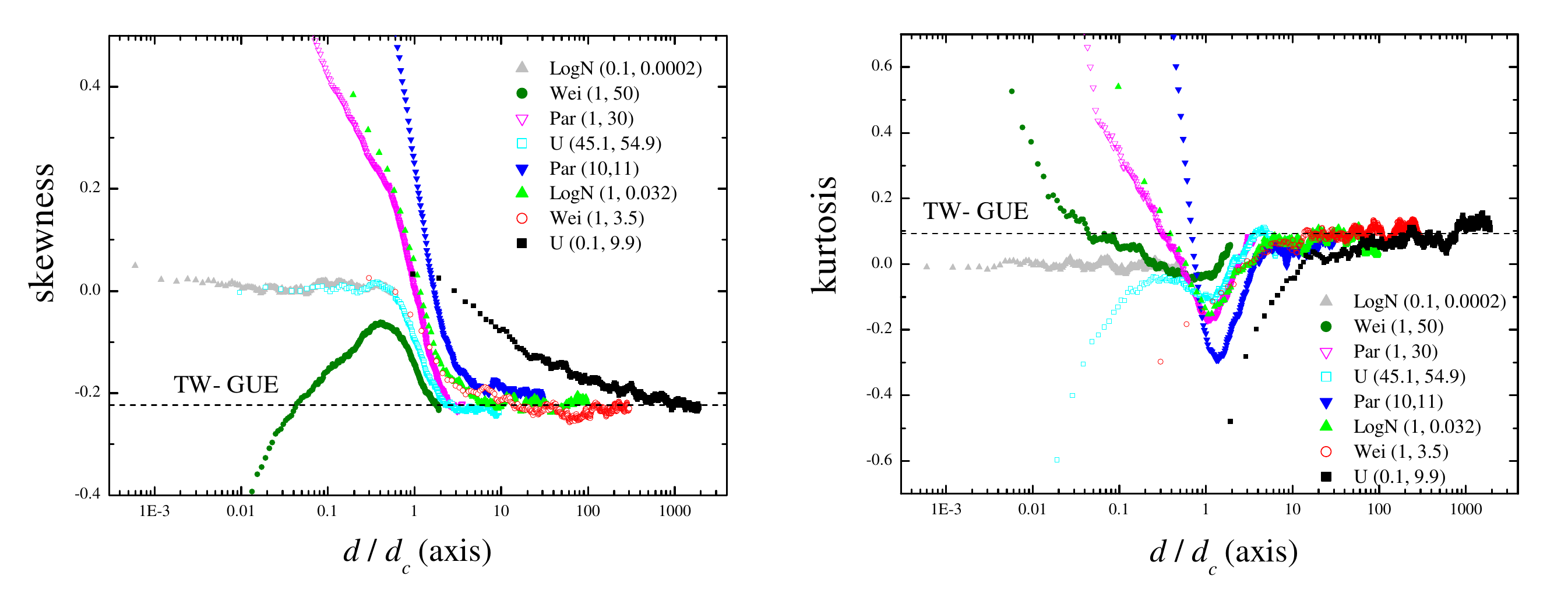}
  \caption{Skewness (left) and kurtosis (right) of the distribution of
    passage times to sites on the axes of the square lattice, as a
    function of the scaled Euclidean distance to the origin for
    different link-time distributions. Results for case U$(0.1,9.9)$
    were obtained for $L=2000$ and from an ensemble of $1.5\cdot10^5$
    points. Horizontal broken lines stand for the TW-GUE values for
    skewness ($-0.224$) and kurtosis ($0.0934$) \cite{Santalla_15}.}
  \label{fig:cumulants_axis}
\end{figure}

\begin{figure} [ht]
  \includegraphics[width=9cm]{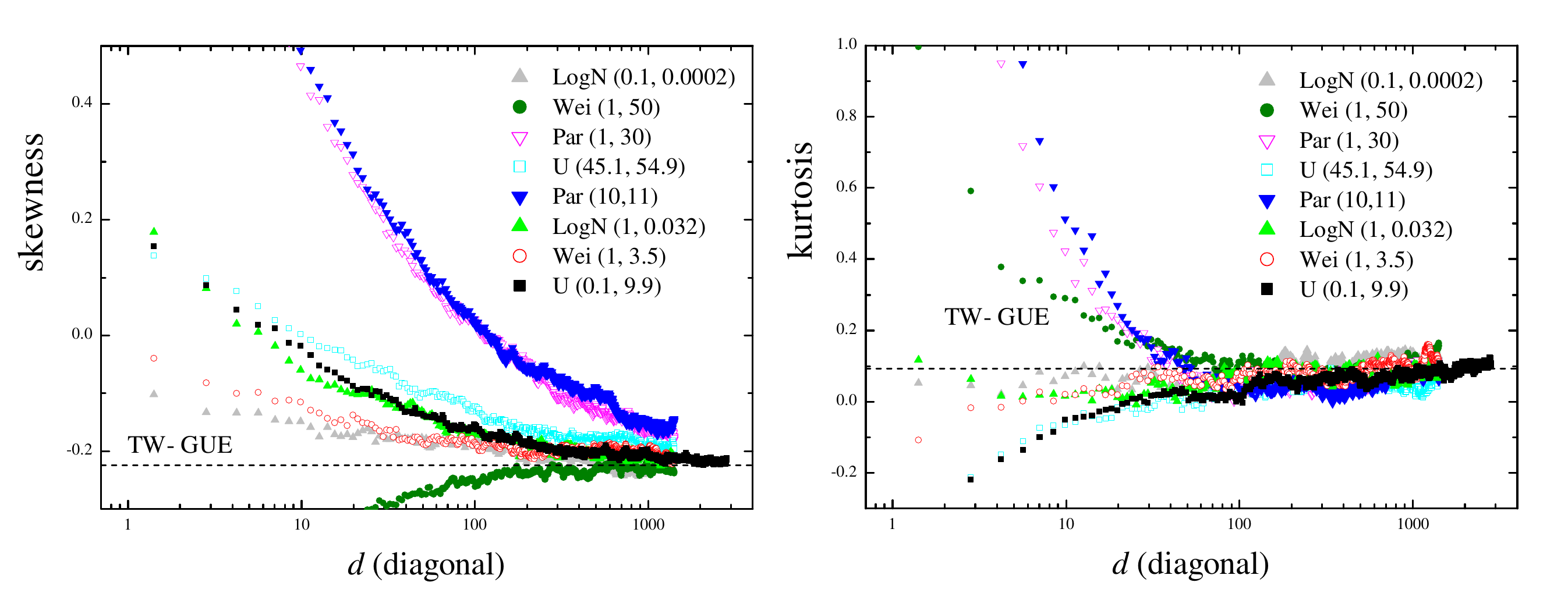}
  \caption{Skewness (left) and kurtosis (right) of the distribution of
    passage times to sites on the diagonal of the square lattice, as a
    function of the bare Euclidean distance to the origin for
    different link-time distributions. Results for case U$(0.1,9.9)$
    were obtained for $L=2000$ and from an ensemble of $1.5\cdot10^5$
    points. Horizontal broken lines stand for the TW-GUE values for
    skewness ($-0.224$) and kurtosis ($0.0934$) \cite{Santalla_15}.}
  \label{fig:cumulants_diag}
\end{figure}

The KPZ class does not only convey the scaling behavior of the
passage-time fluctuations. As discussed in \cite{Santalla_15}, the
full local fluctuation histogram of minimal passage times is predicted
to follow the Tracy-Widom distribution for the Gaussian unitary
ensemble (GUE). This prediction can be checked by measuring higher
order cumulants of the time of arrival distribution, as we have done
in Figs. \ref{fig:cumulants_axis} and \ref{fig:cumulants_diag} for
sites on the axis and the diagonal, respectively. Left parts display
the evolution of the third cumulant (skewness) with the Euclidean
distance to the origin (scaled by $d_c$ in the case of the axis), and
right panels show the results for the fourth cumulant
(kurtosis). Expected TW-GUE values have been represented with
horizontal broken lines.

Results are in agreement with the corresponding behaviors of the passage-time fluctuations discussed above. With regard to the axis, for distances below the crossover length geodesics are straight
lines and the fluctuations in the time of arrival approach the
Gaussian behavior regardless of the type of distribution. Accordingly, for $d \ll d_c$ the cumulants displayed in
Fig. \ref{fig:cumulants_axis} approach or stay close to the Gaussian value of
0. Above $d_c$ a crossover of the fluctuation scaling to the KPZ class
and thus of the cumulants to the TW-GUE values, takes place. With
respect to the diagonal depicted in Fig. (\ref{fig:cumulants_diag}), the curves
monotonically converge to the TW-GUE moments in agreement with the convergence to the KPZ class of the scaling of the passage-time fluctuations displayed in
Fig. \ref{fig:fluct_diag}.

%%%%%%%%%%%%%%%%%%%%%%%%%%%%%%%%%%%%%%%%%%%%%%%%%%%%%%%%%%%%%%%%%%%%%%%%%%%%

\section{Geodesic Deviation}
\label{sec:geodesics}

\begin{figure*}
  \includegraphics[width=\textwidth]{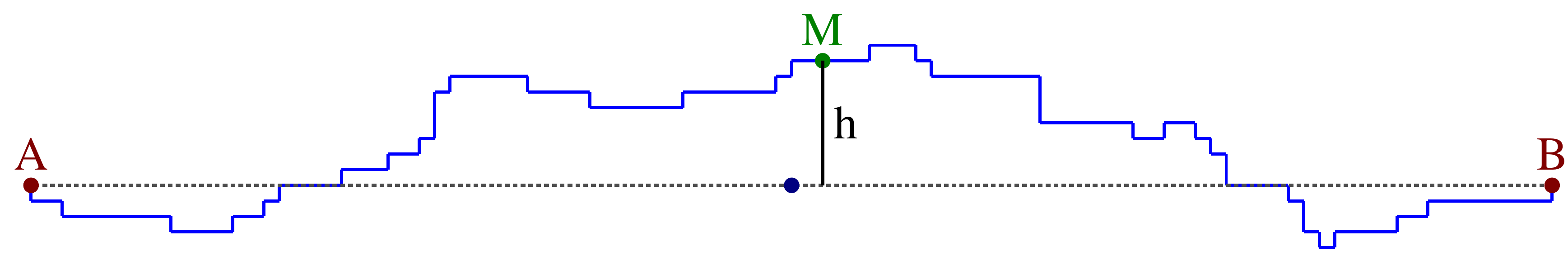}
  \caption{A sample geodesic between two points in the square lattice,
    $A$ and $B$, separated by 100 lattice steps. Point $M$ (marked in
    green) is the middle point, which can be reached from $A$ and $B$
    in the same time. In blue, the middle point of the segment
    $\overline{AB}$, showing that both its $X$ and $Y$ components
    differ from those of $M$.}
  \label{fig:geodesic_illust}.
\end{figure*}

To get a complete characterization of the scaling behavior of the
model we have also focused on a morphological property of the
geodesics, namely, their \emph{lateral deviation}. Let us define the
{\em middle point} of a geodesic as the one reached in the same time
from both extremes, or, in other words, the point reached at half the
total passage time. The lateral deviation of the geodesic, denoted by
$h$, is defined as the Euclidean distance from the middle point to the
straight line joining the endpoints, as illustrated in
Fig. \ref{fig:geodesic_illust}.

\begin{figure} [ht]
  \includegraphics[width=\columnwidth]{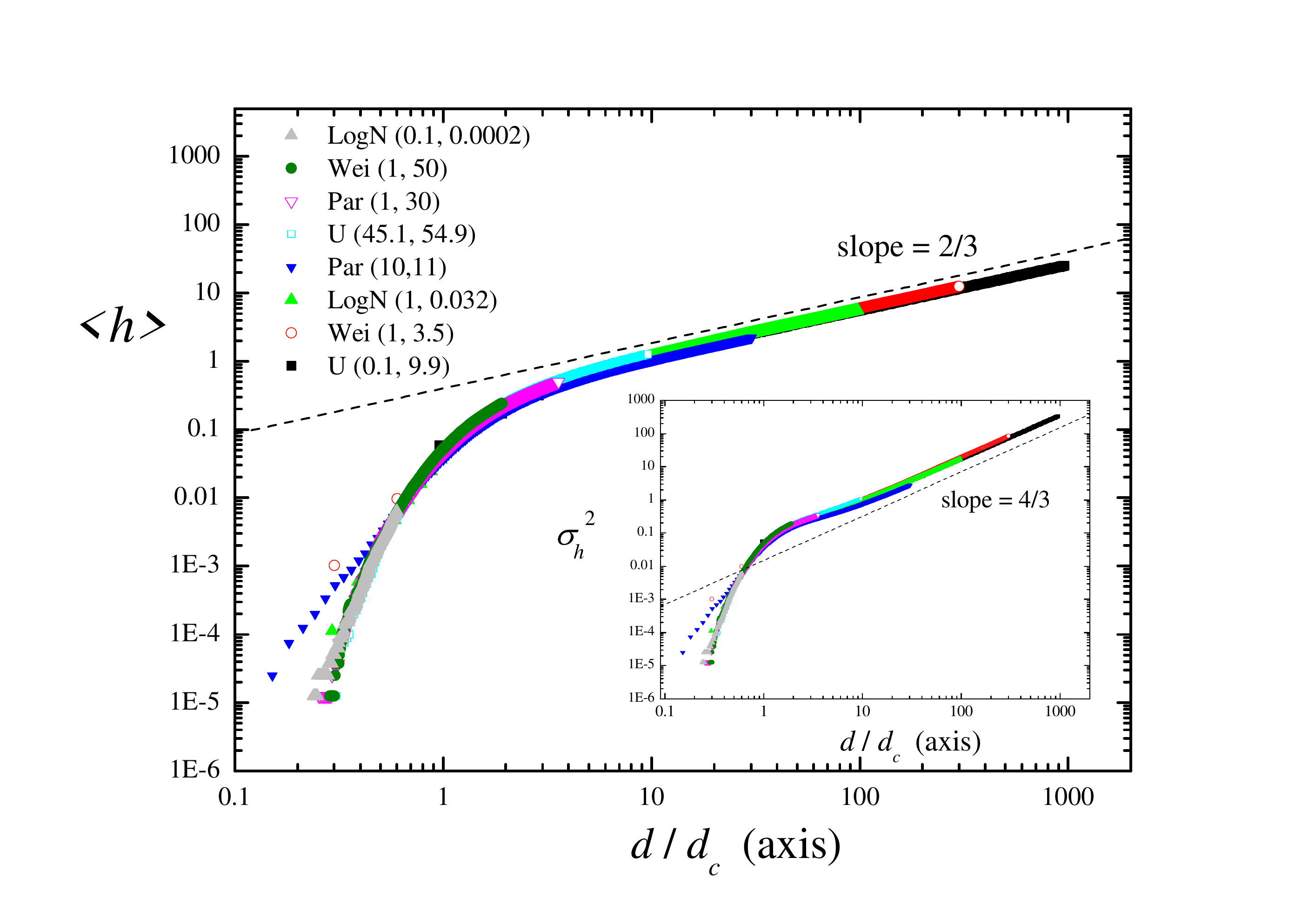}
  \caption{Average lateral geodesic deviation for points on the axis
    as a function of the scaled distance to the origin.(Inset)
    Corresponding variance of the fluctuations. KPZ scaling has been
    represented in both cases with the broken line. }
  \label{fig:lateral_deviation_axis}
\end{figure}

\begin{figure} [ht]
  \includegraphics[width=\columnwidth]{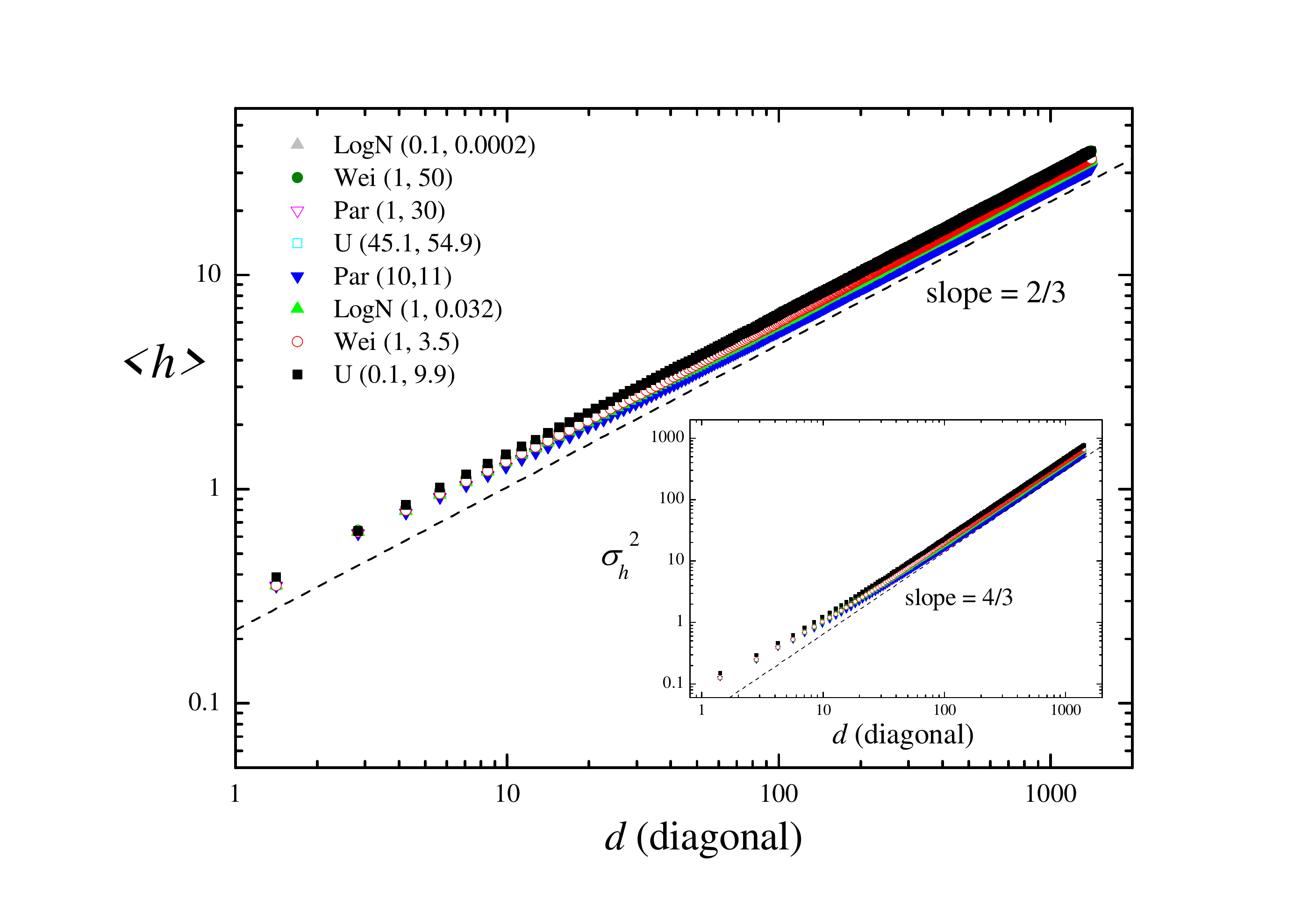}
  \caption{Average lateral geodesic deviation for points on the
    diagonal as a function of the distance to the origin. (Inset)
    Corresponding variance of the fluctuations. KPZ scaling has been
    represented in both cases with the broken line. }
  \label{fig:lateral_deviation_diagonal}
\end{figure}

In a previous work \cite{Santalla_15} it was shown that for random
metrics on 2D manifolds, the lateral deviation of the geodesic scales
with the Euclidean distance $d$ between the points as $h\sim d^{1/z}$,
where $z=3/2$ is the KPZ dynamical exponent. Following our line of
analysis we have computed the average lateral deviation of the
geodesics between the origin and points on the axis and the diagonal,
and the results have been shown in
Figs. \ref{fig:lateral_deviation_axis} and
\ref{fig:lateral_deviation_diagonal} respectively, with the length variance $\sigma^2_h$ displayed in the insets.

In both cases the results are perfectly consistent with our findings
for the fluctuations of the times of arrival. For the geodesics
between points on the axis (Fig. \ref{fig:lateral_deviation_axis}) the
scaling of the lateral deviation has the form:
\beq
h \sim b\(d\over d_c \),
\label{eq:scaling_length_axis}
\eeq
with
\beq
b(x)\sim \begin{cases} 0 & \textrm{if } x\ll 1 \\
  x^{2/3} & \textrm{if } x\gg 1 \end{cases}.
\label{eq:scaling_function_length_axis}
\eeq
As expected, no lateral deviations are observed for $d\ll d_c$ and KPZ
scaling is attained immediately above $d_c$. For the diagonal, the
curves overlap showing a remarkable universal behavior which seems to
be independent on the statistical properties of the local-time
distribution. As for the corresponding passage-time fluctuations,
convergence to the KPZ behavior is very rapid. The same analysis
applies to the behavior of the length variance displayed in the
insets of both figures.

%%%%%%%%%%%%%%%%%%%%%%%%%%%%%%%%%%%%%%%%%%%%%%%%%%%%%%%%%%%%%%%%%%%%%%%%%%%

\section{Limit Shape}
\label{sec:limitshape}

We finish the analysis of the square lattice by addressing the shape
of the geodesic balls $B(t)$ defined in (\ref{eq:balls}). The
\emph{shape theorem} \cite{Richardson_73, Cox-Durrett_81, Kesten_86}
states that $t^{-1}B(t)$ converges in Hausdorff distance as $t\rightarrow\infty$ to a certain non-random, convex, compact set, with
a definite shape which is expected to depend on the distribution of
the passage times between neighboring lattice sites.

In order to characterize this shape we will consider the velocities of growth along the axis and the diagonal. Let
$v_A(d)= \Delta d / \Delta \<T\>(d)$ be the velocity of growth along the
axis at a distance $d$, where $\<T\>(d)$ is the average value of the
minimal time of arrival at that position from the origin. Also, let
$v_D(d)$ the analogous velocity for sites along the diagonal. Note
that we are considering Euclidean distances, not lattice distances, hence
$\Delta d=1$ along the axis and $\Delta d=\sqrt{2}$ along the
diagonal, both in lattice units. We shall also consider the homogeneous case
as a reference. Link-times do not vary but take the uniform value
$\tau$ yielding trivially exact velocities $v_{A0}=\tau^{-1}$ and
$v_{D0}=(\sqrt{2}\tau)^{-1}$, respectively.

\begin{figure} [ht]
  \includegraphics[width=\columnwidth]{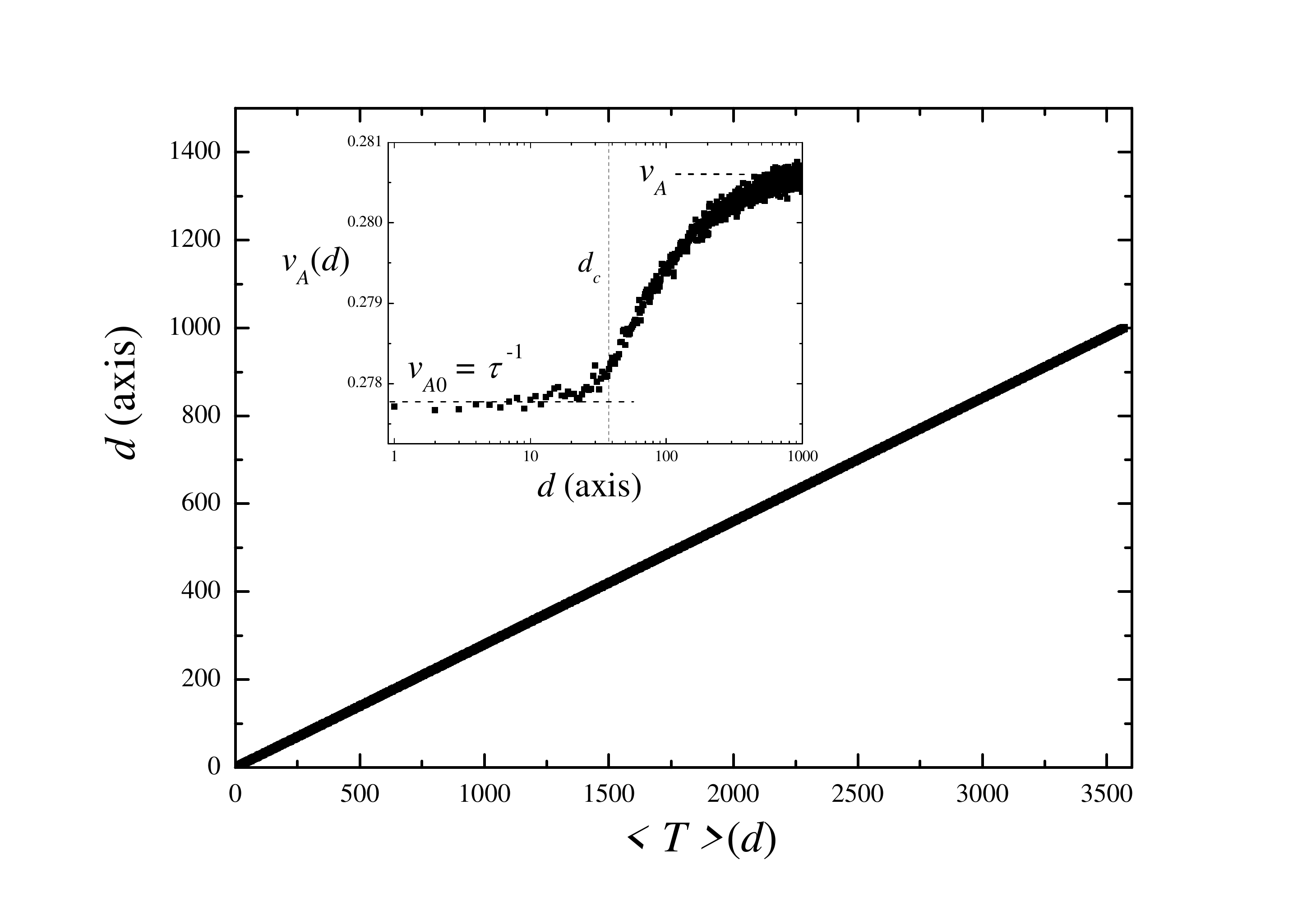}
  \caption{Distance to the origin on the axis direction as a function
    of the average minimal arrival time to reach it. Results
    correspond to case U$(3, 4.2)$. (Inset) Local derivative of the
    data in the main panel, defined as the velocity $v_A(d)$, as a
    function of distance. Horizontal broken lines indicate two
    regimes, $d\ll d_c$, with constant value $v_{A0}$, and $d\gg d_c$,
    with saturation value $v_A$ for the largest $d$ values.}
  \label{fig:velocity_axis}
\end{figure}

\begin{figure} [ht]
  \includegraphics[width=\columnwidth]{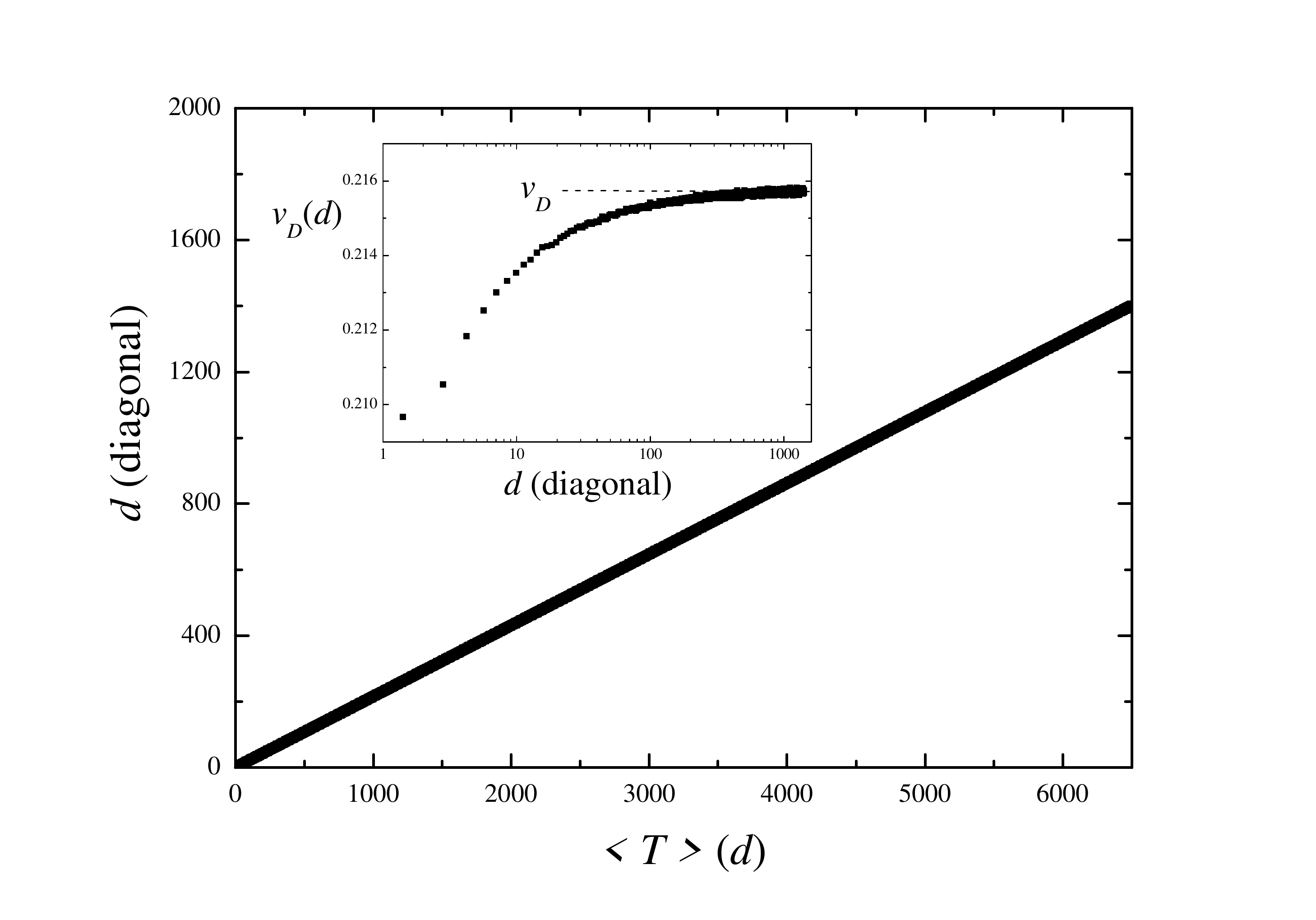}
  \caption{Distance to the origin on the diagonal direction as as
    function of the average minimal arrival time to reach it. Results
    correspond to the U$(3, 4.2)$ link-time distribution. (Inset)
    Local derivative of the data in the main panel, defined as the
    velocity $v_D(d)$, as a function of distance. The horizontal
    broken line indicates the saturation value $v_D$.}
  \label{fig:velocity_diagonal}
\end{figure}

To illustrate the behavior obtained in our model we have displayed in
Figs. \ref{fig:velocity_axis} and \ref{fig:velocity_diagonal} the
results of a representative link-time distribution corresponding to
U$(3, 4.2)$ with $d_c=36$. The figures display the distance $d$ of the
geodesic front along the axis (Fig. \ref{fig:velocity_axis}) and the
diagonal (Fig. \ref{fig:velocity_diagonal}) as a function of the
average minimal arrival time $\< T \> (d)$. Both sets of data display
excellent linear behavior, so that the corresponding velocity of
growth can be accurately estimated from linear regression. However,
when we look at the local derivative displayed in the insets, we
observe a subtle behavior not apparent in the linear plots. With
regard to the growth in the axis direction
(Fig. \ref{fig:velocity_axis}), for points below the crossover
distance $d_c$ the velocity is given by the trivial one $v_{A0}$,
which agrees with the fact that geodesic paths are Euclidean straight
lines. A crossover takes place at $d_c$, beyond which the velocity
increases and stabilizes at a new value which we will call $v_A$ (note
the logarithmic scale for $d$). We can then write:
\begin{equation}\label{eq:vA}
v_A(d)=\left\{
\begin{array}{l l}
            v_{A0} & \mbox{for } d\ll d_c, \\
            v_{A}  & \mbox{for } d\gg d_c.
          \end{array}
\right.
\end{equation}

On the contrary, for growth along the diagonal
(Fig. \ref{fig:velocity_diagonal}), no crossover is observed; just an
initial transient is followed by saturation to a constant value
denoted by $v_D$.

\begin{figure} [ht]
  \includegraphics[width=\columnwidth]{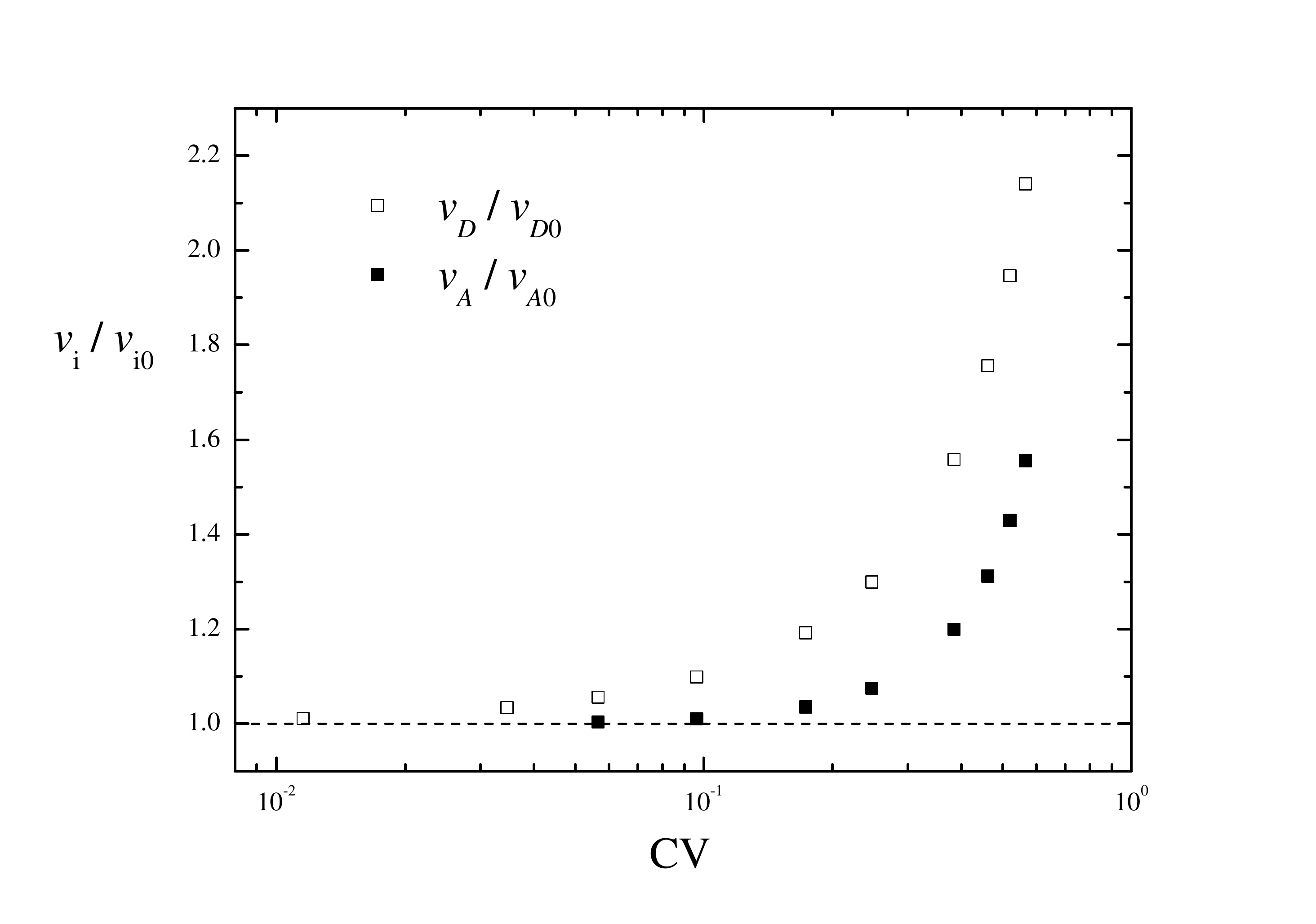}
  \caption{Ratio of the velocities of growth along the axis (solid)
    and diagonal (open) to their homogeneous counterparts as a
    function of the coefficient of variation of the different
    link-time distributions employed in the simulations. Note that
    these results correspond only to uniform link-time distributions.}
  \label{ratio_vi_to_vi0}
\end{figure}

As a consequence of the minimization of the arrival time, the limit
velocities $v_A$ and $v_D$ will always be larger than their uniform
counterparts $v_{A0}$ and $v_{D0}$. However, this effect is more
marked for degenerate directions due to the fact that geodesics do not
need to leave the ensemble of degenerate paths in order to find the
minimal path. We have illustrated this point in
Fig. \ref{ratio_vi_to_vi0} for the uniform distribution with different
parameter values. The limit velocities along the axis and diagonal
have been rescaled by the corresponding clean values, and plotted
against the CV of the distribution. In all cases the increase of the
velocity is larger for the diagonal.

Another remarkable result is that these ratios are unambiguously
determined by the CV, i.e. distributions with different parameter
values but the same CV yield the same values for $v_A / v_{A0} $ and
$v_D / v_{D0}$, so that they are undistinguishable in the figure. It
should be noticed that when non-uniform distributions are used for the
link-times, the results remain qualitatively only. A similar
collapse to a single curve as in Fig. \ref{ratio_vi_to_vi0} is only
obtained when the same type of distribution is used, changing its
parameters. Interestingly, the two relative velocities increase with
CV in a monotonic way, with $v_A \rightarrow v_{A0}$ and $v_D
\rightarrow v_{D0}$ as $\mbox{CV} \rightarrow 0$. This is consistent
with the fact that at this limit the distribution behaves as the Dirac
delta function $\delta(t-\tau)$ and the trivial homogeneous case is
recovered.

\begin{figure} [ht]
  \centering
  \includegraphics[width=\columnwidth]{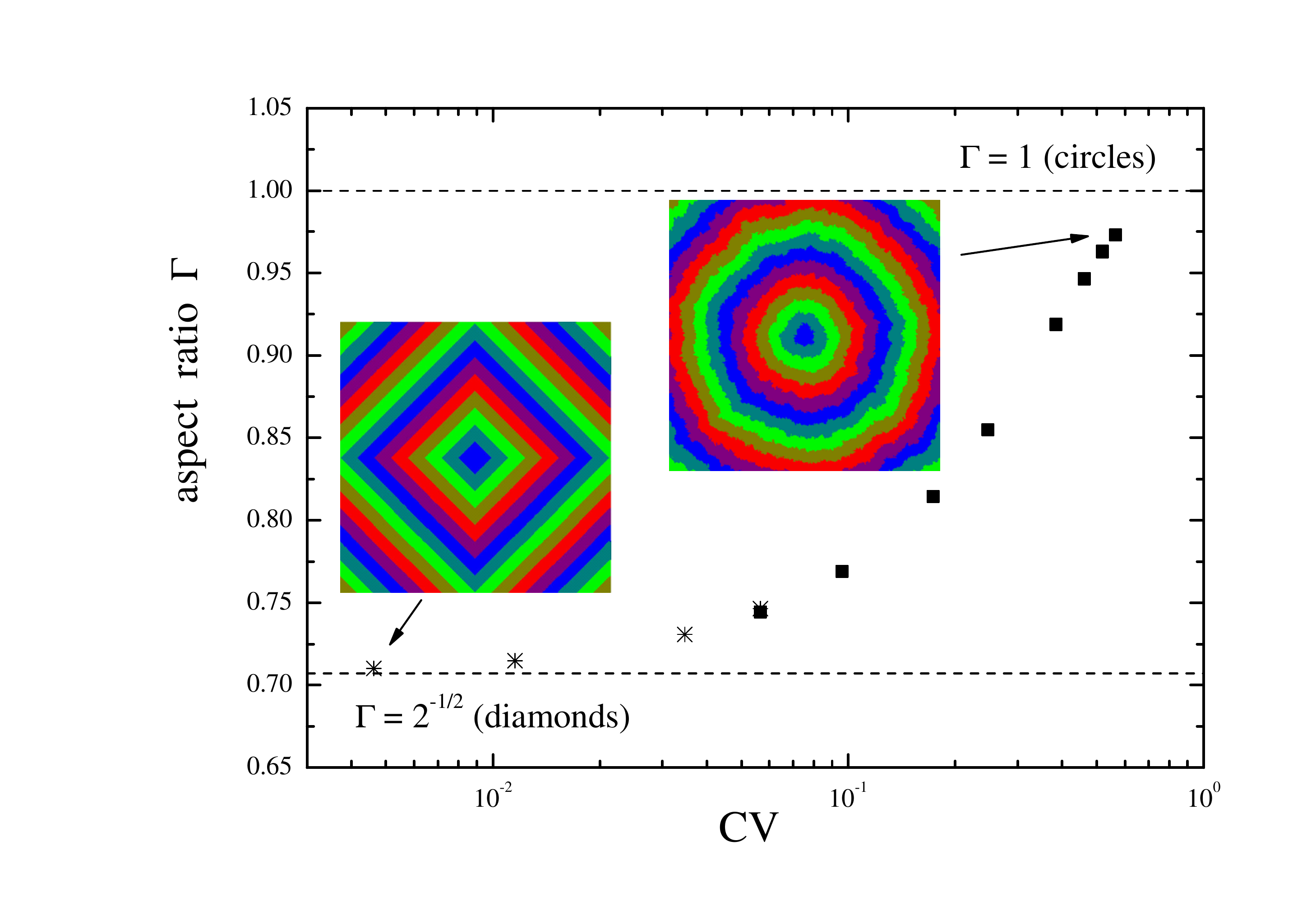}
  \caption{Aspect ratio of the geodesic balls, defined in
    Eq. (\ref{eq:aspect_ratio}), as a function of the coefficient of
    variation for the same results displayed in
    Fig. \ref{ratio_vi_to_vi0}. Crosses stand for those values
    obtained after the $v_A=v_{A0}$ approximation. Limiting cases have
    been indicated with horizontal broken lines. Illustrations depict
    the growth of the balls $B(t)$ in a $401\times401$ lattice for the
    two extreme cases for the link-time distribution: (left) U$(4.96,
    5.04)$ with color changing after $\Delta t=125$; (right) U$(0.1,
    9.9)$ with $\Delta t=60$.}
  \label{fig:aspect_ratio}
\end{figure}

It should be mentioned that the lattice size $L$ establishes a lower bound for the
value of CV that allows determining the limit velocity
$v_A$. Indeed, as shown in Eq. \eqref{eq:vA}, this limit velocity is
attained when $d\gg d_c$ or, from Eq. \eqref{eq:def_dc}, when $d\gg
3^{-1}\mbox{CV}^{-2}$. On the other hand, all properly measured
distances should be smaller than the system size, $d\ll L$. This
results in $\mbox{CV} \gg (3L)^{-1/2}$, thereby establishing a lower
bound for the possible values of CV that allow for a reliable measurement
of $v_A$. For example, for $L=1000$ we have $\mbox{CV} \gg
0.018$. When $\mbox{CV} \ll (3L)^{-1/2}$ (or $d_c \gg L$), the velocity
along the axis will then be given by $v_{A0}$. At the other extreme,
the CV is bounded above by $3^{-1/2}$, which is an intrinsic property
of the uniform distribution. This results in the range $[0.018,0.577]$ for the available values of CV in a lattice with $L=1000$ and uniformly distributed link times.
We then define the aspect ratio $\Gamma$ of the geodesic balls as the
ratio between the two limit velocities:
\begin{equation}
  \Gamma \equiv \frac{v_D}{v_A}.
  \label{eq:aspect_ratio}
\end{equation}
Results for the aspect ratio have been displayed in
Fig. \ref{fig:aspect_ratio} as a function of the coefficient of
variation. For those values of CV not satisfying the $\mbox{CV} \gg
0.018$ condition discussed above, we have considered $v_{A0}$ instead
of $v_A$ (crosses in the figure). Despite being a transient regime, we
see in Fig. \ref{ratio_vi_to_vi0} that $v_A \rightarrow v_{A0}$ as
$\mbox{CV} \rightarrow 0$, hence we can then assume that the
difference between $v_A$ and $v_{A0}$ is negligible when $\mbox{CV}
\ll 0.018$. This approximation is validated by the continuity of the
points displayed in the figure at $\textrm{CV} \approx 0.018$.

The aspect ratio of the geodesic balls is again completely determined
by the coefficient of variation. As CV increases, the shape evolves from
the diamond structure given by $\Gamma = (\sqrt{2})^{-1}$, attained at the
limit $\mbox{CV} \rightarrow 0$ (homogeneous case), towards the circular contour given by $\Gamma = 1$. These two limit shapes have been illustrated with the balls obtained at the extreme values of CV.

%%%%%%%%%%%%%%%%%%%%%%%%%%%%%%%%%%%%%%%%%%%%%%%%%%%%%%%%%%%%%%%%%%%%%%%%%%%%

\section{Delaunay Lattices}
\label{sec:delaunay}

The previous sections have discussed the FPP model on a square lattice. In
such regular systems it is quite straightforward to recognize both unique
and degenerate directions, as illustrated in Fig. \ref{fig:illust_degeneracy}. Also, it is rather easy to calculate the exact geodesic degeneracy for each lattice direction, as we did in Eq. \eqref{eq:number_geodesics_homoheneous} for the square lattice. It is therefore very pertinent to ask whether non-regular lattices might lead to different behavior. It seems reasonable to think that disordered lattices in general will present a certain degree of geodesic degeneracy that, contrary to ordered lattices, will be isotropic and dependent only on distance between the nodes. This degeneracy will increase with $d$ and will be subject to some fluctuations.

In this section we consider the FPP model on disordered planar lattices built as {\em Delaunay lattices}. A Delaunay lattice is a triangulation which fulfills a certain optimality condition: the circumscribed circle built on any triangle does not contain any other lattice points. Given a set of $N$ points on the plane, the Bowyer-Watson algorithm \cite{Bowyer_81,Watson_81} builds a Delaunay lattice in $O(N\log(N))$ steps in average, or $O(N^2)$ in the worst cases \cite{Rebay_93}.

\begin{figure}[ht]
  \centering
  \includegraphics[width=\columnwidth]{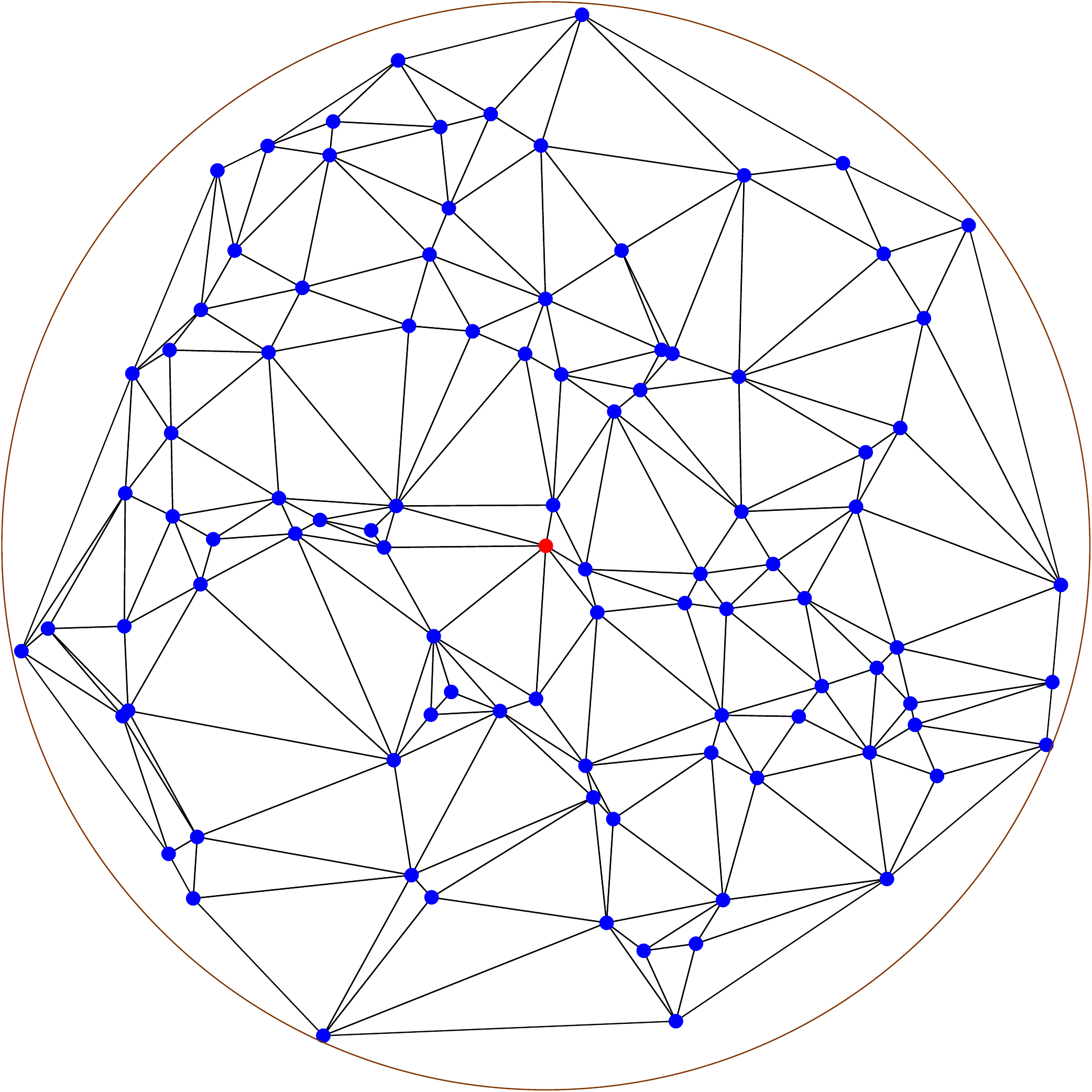}
  \caption{Example of Delaunay lattice built on $N=100$ points
    randomly distributed on the unit circle. The point marked in red
    is at the center.}
  \label{fig:delaunay}
\end{figure}

The geometric setup is as follows. We consider the unit circle with
the central node $\textbf{x}_0$ at its geometrical center. Then we
mark a set of $N_m$ points at fixed distances from the center, where
measurements will take place. To avoid unwanted correlations,
measurement points are homogeneously distributed on a spiral so that
the coordinates of the $j$-th point $\textbf{x}_j$ ($j=1,\dots,N_m$)
are $x_j=j/N_m\sin(j2\pi\varphi)$ and $y_j=j/N_m\cos(j2\pi\varphi)$,
where $\varphi$ is the golden ratio. Accordingly, the Euclidean
distance of these points to the origin is
$d(\textbf{x}_j)=j/N_m$. Next, we choose other $N-N_m$ uniformly
distributed random points on the circle and we build the Delaunay
lattice of the whole set of points (see Fig. \ref{fig:delaunay} for an
example). Notice that the lengths of the resulting links can vary
notably. As for the square lattice, we associate to each link a
crossing time, which is randomly chosen from a given probability
distribution, hence disregarding the actual length of the
link. Finally, we obtain the minimal traveling time from the origin to
all lattice points.

\FloatBarrier

Each simulation of the system corresponds to a different realization of the link time distribution, always using the same fixed lattice. It must be stressed that simulations of Delaunay lattices are more demanding computationally than for the square lattice, so the ensemble of realizations performed here is markedly less significant. Besides, since qualitative behavior does not depend on the link-time distribution function, for simplicity we will only consider the uniform distribution. The presentation of the results will follow the same scheme as used for the square lattice.

\begin{figure} [ht]
  \centering
  \includegraphics[width=\columnwidth]{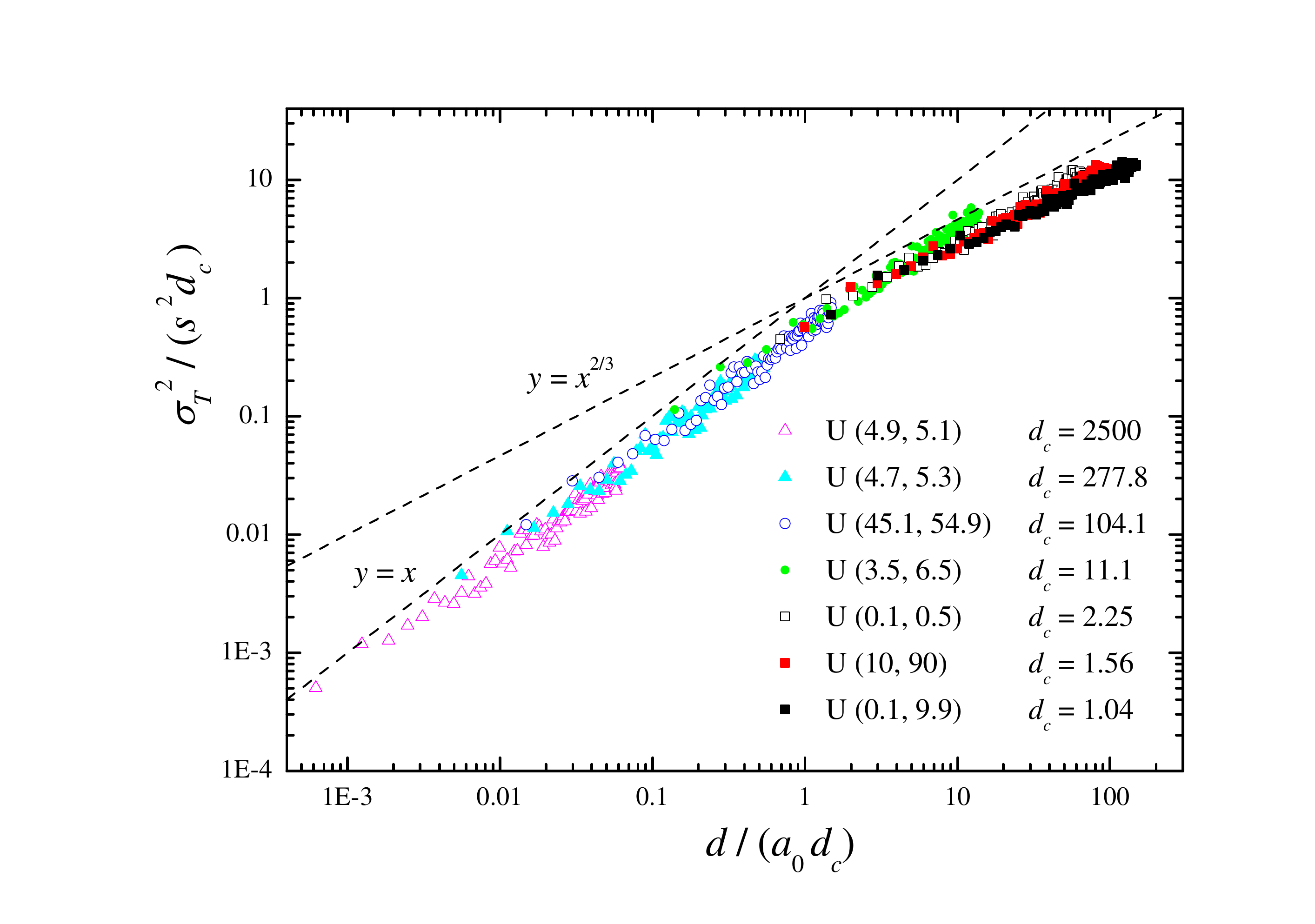}
  \caption{Scaled variance of the arrival time as a function of the
    Euclidean distance to the center, scaled with the mean link length
    $a_0$ and $d_c$, in the Delaunay lattice for different parameters
    of the uniform link-time distribution. Results were obtained from
    $1000$ simulations and $N=10^5$. Broken lines indicate the two
    branches of the scaling function given in Eq.
    (\ref{eq:scaling_function_axis}).}
  \label{fig:fluct_time_delaunay_quenched}
\end{figure}

We start by showing in Figure \ref{fig:fluct_time_delaunay_quenched}
the fluctuations in the minimal arrival time as a function of the
distance to the center. As in Fig. \ref{fig:fluct_axis} (right), the
variance of the passage time has been expressed in units of $s^2 d_c$,
while the Euclidean distance to the center, $d$, has been firstly
scaled in terms of lattice jumps by a certain characteristic lattice
length $a_0$, and then rescaled by the crossover distance
$d_c$. Different values for $a_0$ have been employed, all leading to
very similar results. Hereafter we will consider that the
characteristic length $a_0$ is given by the mean link length obtained
from the link length distribution, which for the fixed lattice
employed here ($N=10^5$) was $0.0064$ with a standard deviation of
$0.0033$. However, results do not change significantly if we consider
the average geometric distance between nodes calculated as
$\sqrt{\pi/N}$ (equal to $0.0056$ for $N=10^5$), or whether, instead
of $d/a_0$, we consider the exact number of links included in the
minimal path of the homogeneous case $s^2=0$, which we will denote by
$n_H(d)$.

\begin{figure} [ht]
  \centering
  \includegraphics[width=\columnwidth]{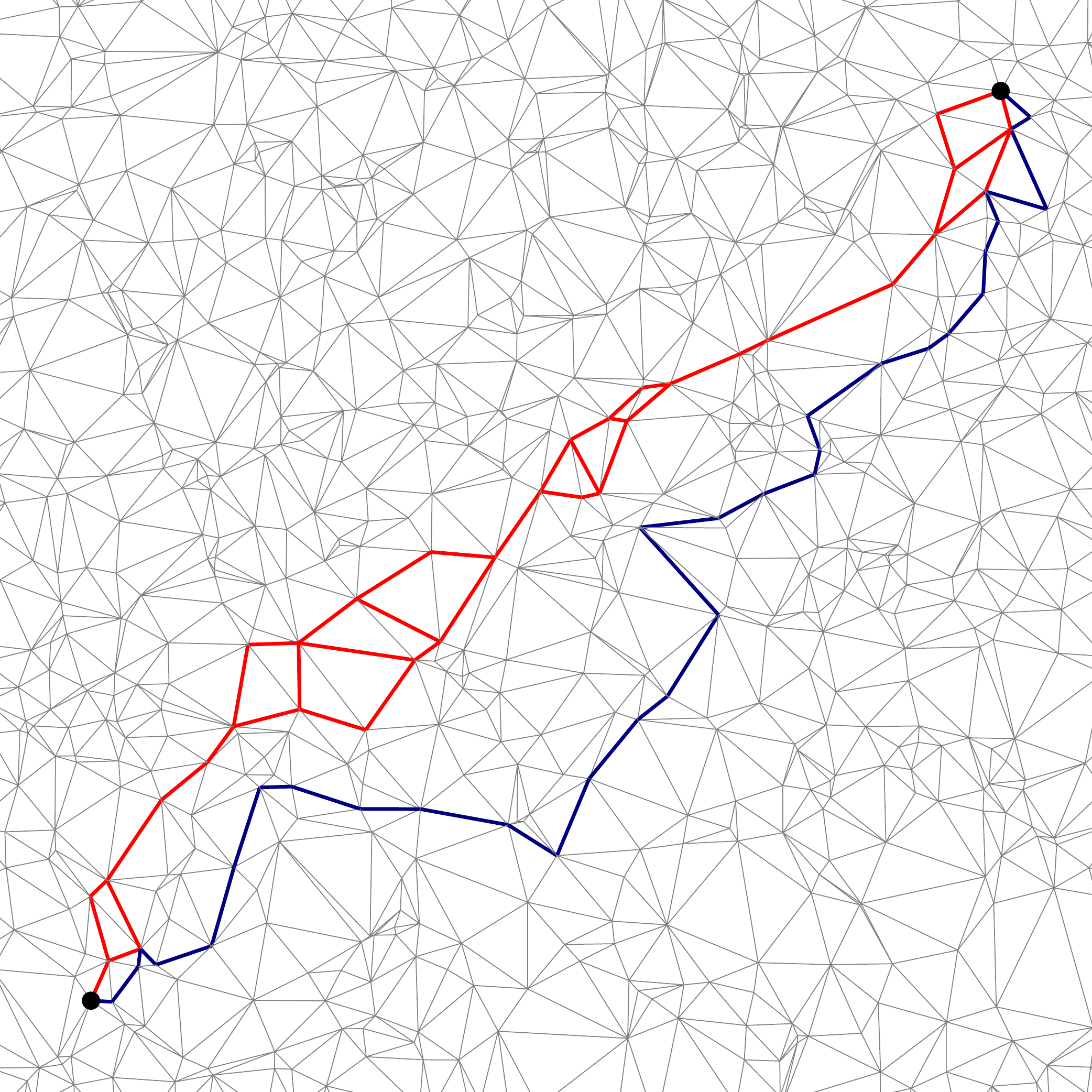}
  \caption{Geodesic degeneracy in Delaunay lattices. Red color highlights the ensemble of minimal paths obtained in the homogeneous $s^2=0$ case between the two lattice nodes identified by the thick points. We have $n_H=23$. Dark blue color identifies a geodesic path obtained in a given realization of the case U(0.5,2), which is made up of $n=33$ links.}
  \label{fig:geodesic_degeneracy_delaunay}
\end{figure}

Results shown in Fig. \ref{fig:fluct_time_delaunay_quenched} for the
Delaunay lattice are quite similar to those displayed in
Fig. \ref{fig:fluct_axis} for the axis direction on the square
lattice. A fairly good collapse to the scaling function given in
Eq. \eqref{eq:scaling_function_axis} is observed. This result allows
us to claim that the effective geodesic degeneracy in the Delaunay
lattice is rather weak, although it is not exactly null as it was
along the axis on the square lattice. Indeed, as we illustrate in
Fig. \ref{fig:geodesic_degeneracy_delaunay}, the minimal path between
two nodes in the homogeneous case may experience local bifurcations
without this entailing additional links. This contributes to the
geodesic degeneracy in terms of $N_{deg}$ (which also increases
exponentially) but has little impact on the results because the degree
of overlap among the degenerate geodesics is very significant (notice
that there are many links which are shared by all the paths), contrary
to what happens along the diagonal direction on the square lattice,
for instance. This behavior highlights the need to refine our
definition of geodesic degeneracy, which so far has been based on
the number $N_{deg}$ of degenerate paths. Although it can be
considered as a first approximation, it clearly does not convey any
information on the overlap among the different paths, which is
certainly relevant and will be a matter of ongoing work.

\begin{figure} [ht]
  \centering
\includegraphics[width=\columnwidth]{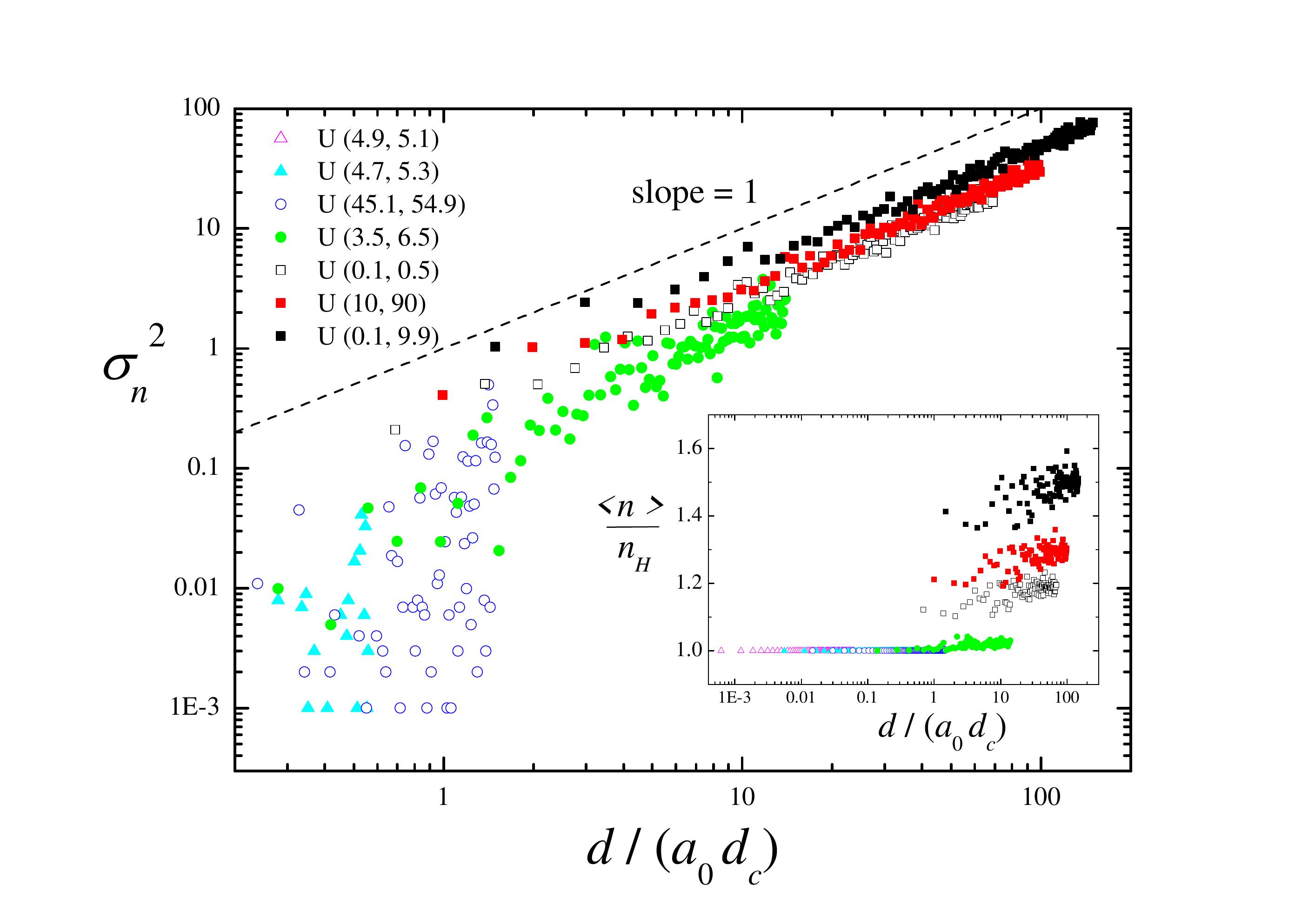}
    \caption{Variance of the number of links of the minimal path as a function of the scaled distance in the Delaunay lattice for the same set of link-time distributions displayed in Fig. \ref{fig:fluct_time_delaunay_quenched}. The broken line indicates the expected linear behavior. (Inset) Corresponding average number of links in the geodesic path divided by the number of links in the minimal path of the homogeneous $s^2=0$ case, $n_H$.}
  \label{fig:fluct_number_steps_delaunay_quenched}
\end{figure}

\begin{figure} [ht]
  \centering
 \includegraphics[width=\columnwidth]{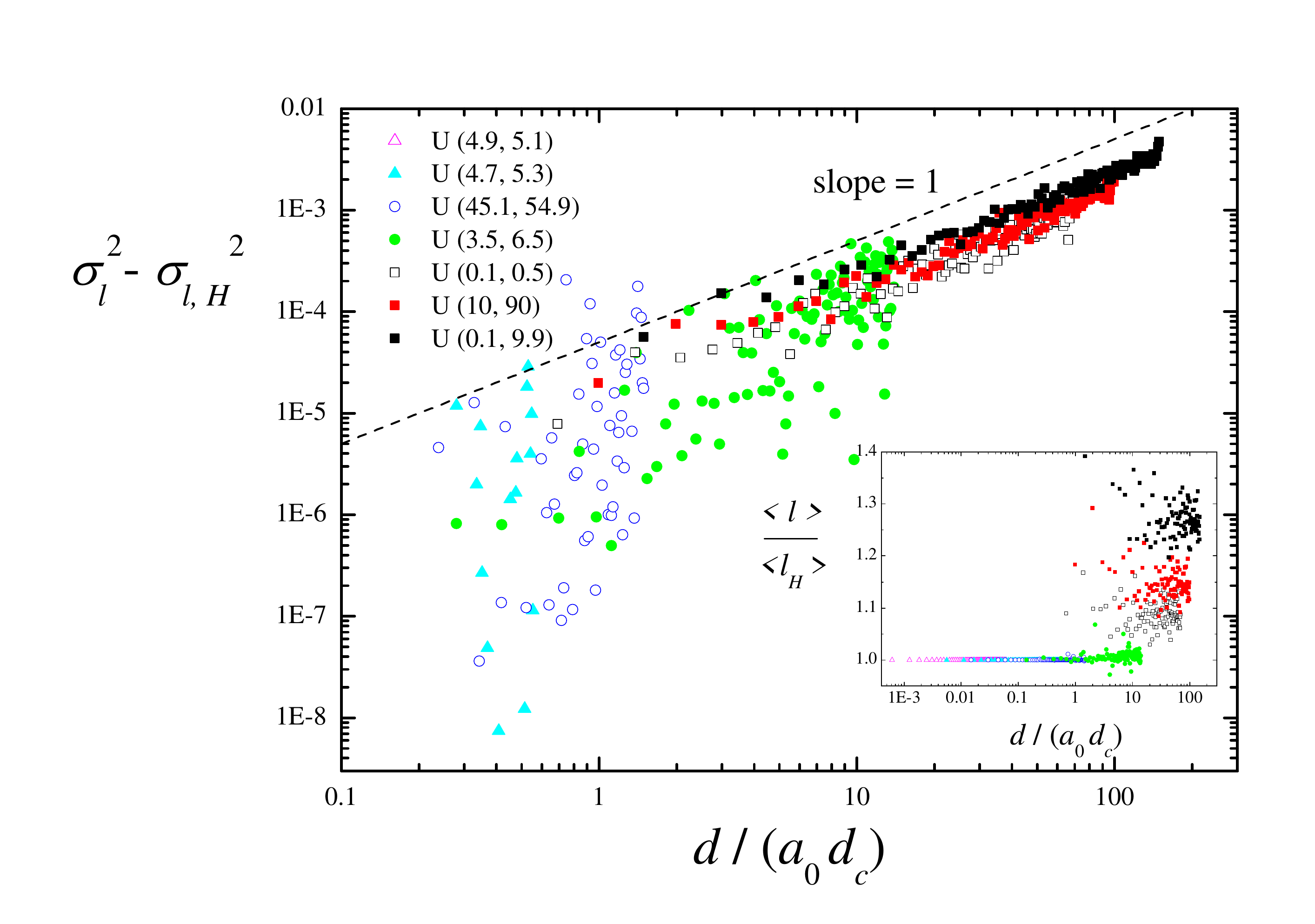}
    \caption{Variance of the length of the minimal path, corrected by the variance of the geodesic length in the homogeneous $s^2=0$ case, as a function of the scaled distance in the Delaunay lattice for the same set of results displayed in Fig. \ref{fig:fluct_time_delaunay_quenched}. The broken line indicates the expected linear behavior. (Inset) Corresponding average path length divided by its homogeneous counterpart, $\langle l_H \rangle$. }
  \label{fig:fluct_length_delaunay_quenched}
\end{figure}

The consequences of the weak geodesic degeneracy in Delaunay lattices
are also reflected in the geometrical properties of the minimal path
such as the geodesic length $l$. In the square lattice this length was
uniquely determined by the number $n$ of links that make up the
geodesic path, see Eq. \eqref{eq:geodesic_lenght}. For disordered
lattices, however, there is a distribution of link lengths so that
these two magnitudes must be considered separately. Figure
\ref{fig:fluct_number_steps_delaunay_quenched} displays the variance
of the number of steps involved in the minimal path, $\sigma^2_n$, as
a function of the rescaled distance. The inset shows the expected
value of that magnitude, $\<n\>$, divided by the value obtained for
$s^2=0$, $n_H$. Fig. \ref{fig:fluct_length_delaunay_quenched} shows
corresponding results for the actual geodesic length, $l$, which hence
takes into account actual link lengths. In the main part we display
the length variance, $\sigma_l^2$, corrected by subtracting the length
variance of the homogeneous $s^2=0$ case, denoted as $\sigma_{l,
  H}^2$, which can be viewed as a sort of \emph{intrinsic variance}
associated to the lattice disorder. In the inset we show the average
geodesic length, $\<l\>$, divided by the same magnitude in the $s^2=0$
case, denoted by $\langle l_H \rangle$.

We must first note the similarity between both plots despite the fact
that the actual length involves geometrical aspects of the lattice not
considered in the number of links. Both Figures
\ref{fig:fluct_number_steps_delaunay_quenched} and
\ref{fig:fluct_length_delaunay_quenched} show a reasonably good data
collapse, similar to that found in Fig. \ref{fig:fluct_length_axis}
for the geodesic length along the axis direction in the square
lattice, adding consistency to our claims. There, the geodesic for the
clean case was unique and trivial. Here, the degeneracy of the minimal
path in the clean case contributes with an intrinsic dispersion that
must be corrected in the raw data to recover the expected
behavior. Note also the large fluctuations obtained when
$d/(a_0d_c)<1$, which are a consequence of the intrinsic disorder in
the lattice topology. For Delaunay lattices, the cost of a deviation
from the geodesic of the homogeneous case is smaller than for the axis
on the square lattice. Besides, the remaining geodesic degeneracy for
$s^2=0$ also favors deviations by enlarging the space of possible
optimal paths. Therefore, the expected fluctuations for $d/a_0<d_c$
are larger than on the square lattice, justifying the scattered data
for low values of $d$ in both figures.

\begin{figure} [ht]
  \centering
\includegraphics[width=\columnwidth]{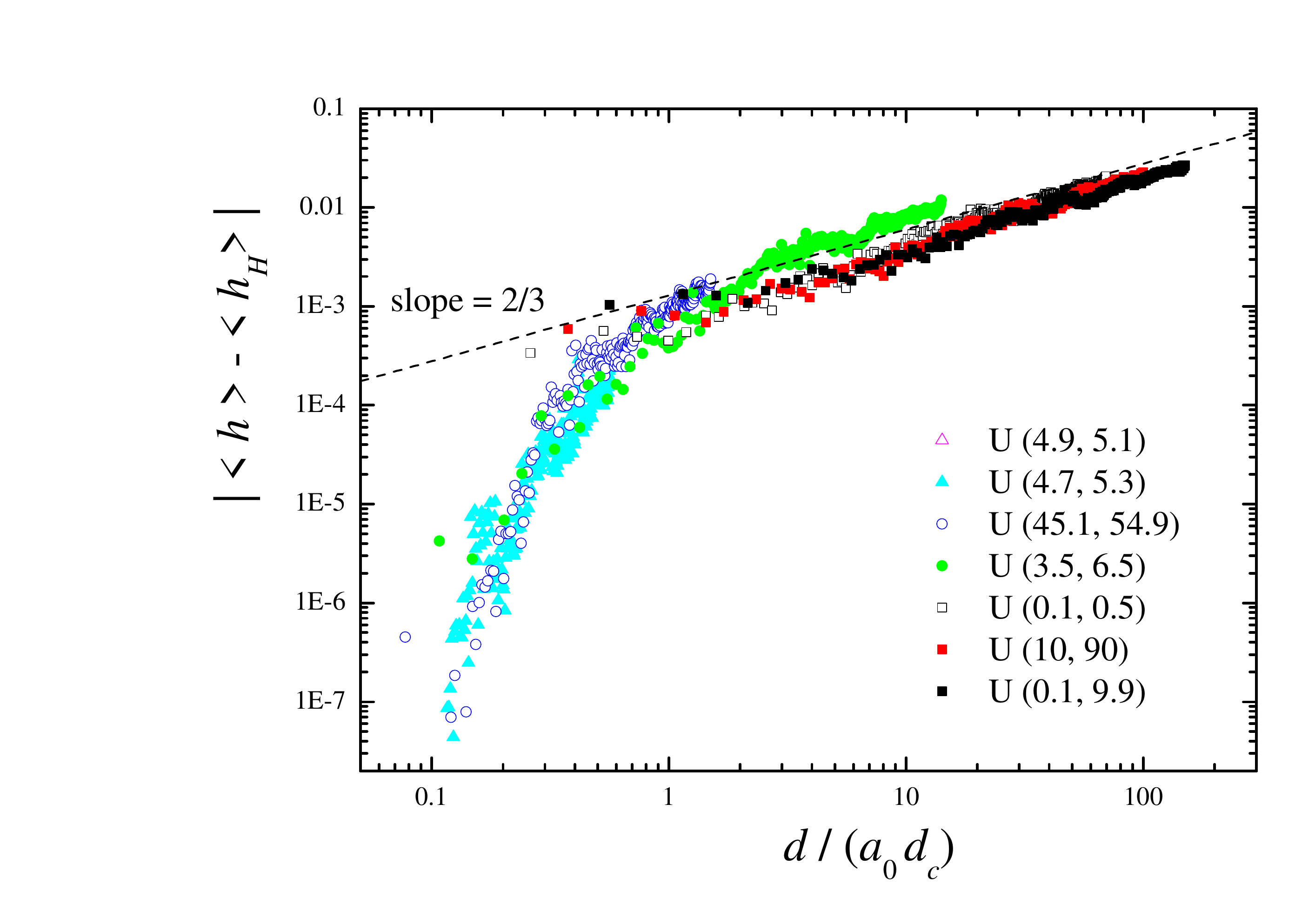}
      \caption{Average lateral deviation corrected by the value of the
        homogeneous case as a function of the scaled distance in the
        Delaunay lattice. KPZ scaling has been represented with the
        broken line. Results for a given distance $d$ were obtained
        from the average over all lattice nodes in a circular ring
        centered at the origin with radius $d$ and thickness
        $dr=1/300$.}
  \label{fig:lateral_deviation_delaunay}.
\end{figure}

We finally address the lateral deviation of the geodesics in Delaunay lattices. As discussed in Sec. \ref{sec:geodesics}, the lateral deviation was defined as the Euclidean distance from the half-time point of the geodesic to the straight line joining the end nodes. For the square lattice no further distinction was necessary inasmuch as that line had a clear physical meaning: in the case of the axis direction it accounted for the geodesic path of the homogeneous system, while for the diagonal direction it represents the average path of the ensemble of degenerate geodesics. In Delaunay lattices, however, the average geodesic path of the clean case is no longer a straight line (see Fig. \ref{fig:geodesic_degeneracy_delaunay}) but an intricate curve to which we can associate a mean lateral deviation $\< h_H(d)\>$  obtained from the average of the lateral distance of the degenerate (or not) middle point. Note that we are considering absolute values for $h$.
Following our line of argumentation, we must correct the raw average value obtained from different realizations of the link-time distribution, $\langle h(d)\rangle$, by subtracting the average deviation of the clean case. The behavior of the resulting corrected lateral deviation $|\< h(d)\>-\< h_H(d)\>|$ has been displayed in Fig. \ref{fig:lateral_deviation_delaunay}. We must point out that preliminary results obtained at the fixed measurement points were rather noisy, so we have performed an average of this quantity over all lattice points at distance $(d, d+dr)$. We can readily recognize in Fig. \ref{fig:geodesic_degeneracy_delaunay} the behavior displayed in Fig. \ref{fig:lateral_deviation_axis} for the axis direction in the square lattice.

%The variance of $ \langle | h-\langle h_H\rangle | \rangle$ is:
%\begin{align*}
%  \sigma^2(| h-\langle h_H\rangle |) & =  \langle (h-\langle h_H\rangle)^2  \rangle -  \langle | h-\langle h_H\rangle |  \rangle^2 \\
%   & =  \langle h^2 \rangle -2 \langle h \rangle  \langle h_H \rangle +  \langle h_H \rangle ^2-\langle | h-\langle h_H\rangle |  \rangle^2 \\
%   & =  \sigma^2_h + \langle h \rangle^2 -2 \langle h \rangle  \langle h_H \rangle +  \langle h_H \rangle ^2-\langle | h-\langle h_H\rangle |  \rangle^2
%\end{align*}

Our claim that the geodesics in a fixed Delaunay lattice present low
degeneracy receives further support from a probabilistic argument. Let
$P(h)$ be the probability density function for the lateral deviation
of all geodesics joining two fixed points. For strong disorder in the
link-times, the geodesics ensemble form a wide cloud and the
distribution $P(h)$ is very broad. In the homogeneous limit, the
distribution must be narrower. Yet, the variance is sometimes high
because the geodesic ensemble reduces to a few curves which can be
very distant. Thus, $P(h)$ in the homogeneous limit is conformed by a
sum of delta peaks, and the variance is not a good measure of its
concentration. A better observable is given by the {\em entropy} of
the distribution \cite{Vasicek.76}, specifically the Kullback-Leibler
divergence \cite{Desurvire_09} which quantifies the relative entropy
with respect to a homogeneous probability distribution on a fixed
interval \cite{note_entropy}.

\begin{figure}
\includegraphics[width=8cm]{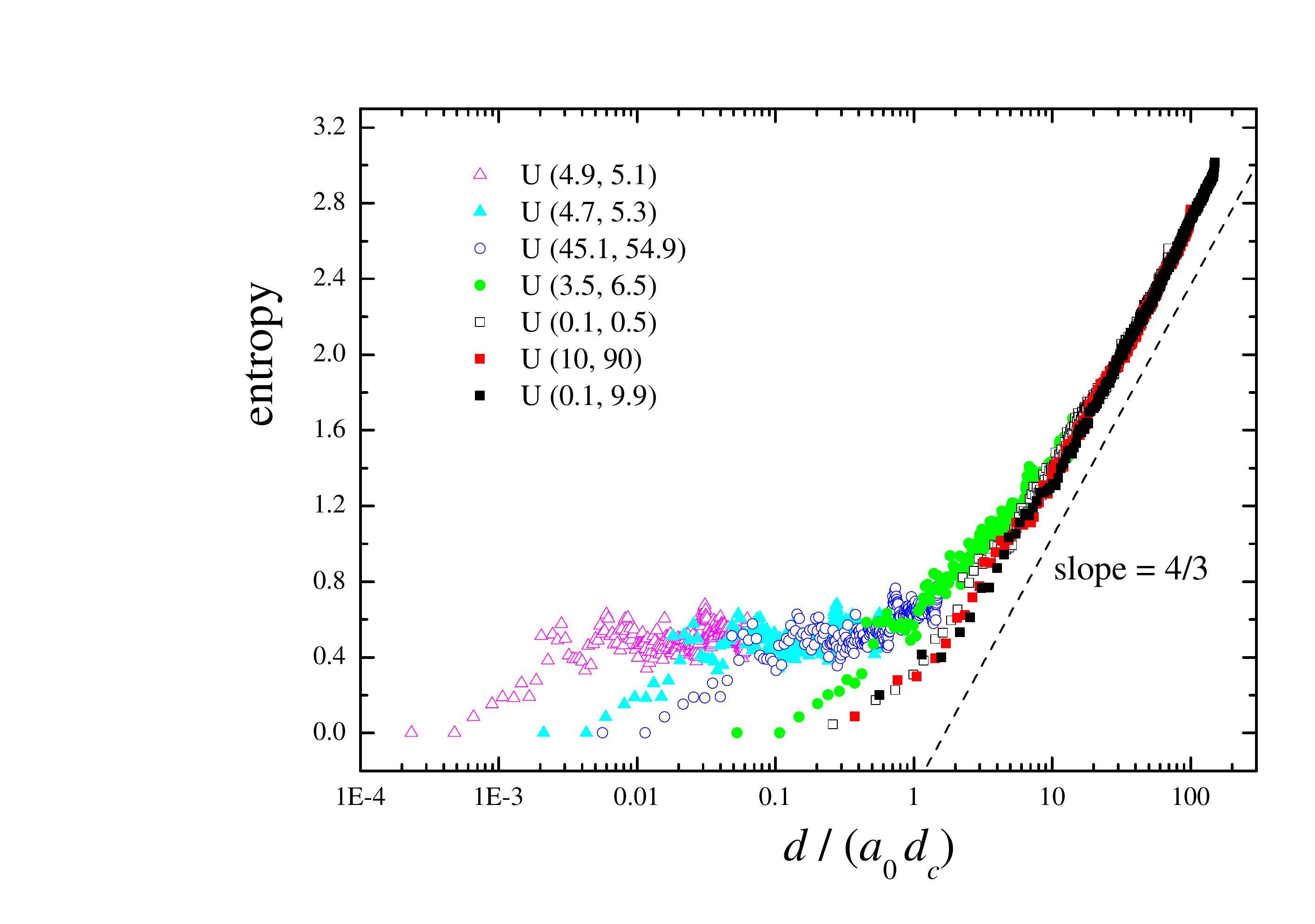}
\caption{Relative entropy of the probability density function for
  lateral deviations of the geodesic ensemble as a function of the
  rescaled Euclidean distance in the Delaunay lattice for the same
  system displayed in
  Fig. \ref{fig:lateral_deviation_delaunay}. Notice the logarithmic
  scale on the horizontal axis.}
\label{fig:entropy}
\end{figure}

Figure \ref{fig:entropy} depicts such an average relative entropy of
$P(h)$ for the geodesic ensembles previously analysed, as a function
of the Euclidean distance between the extreme points in units of $(a_0
d_c)$. For very low $d/(a_0d_c)$ the entropy approaches a fixed value,
related to the expected number of different geodesic paths in the
homogeneous Delaunay lattice. For $d/(a_0 d_c)\gg 1$, the collapse of
the curves is remarkable; the entropy grows logarithmically with the
distance, making the curve appear as a straight line due to the
logarithmic scale chosen for the horizontal axis. Since the relative
entropy of a Gaussian distribution with variance $\sigma^2$ is
proportional to $\log(\sigma^2)$, we may conjecture that the entropy
of $P(h)$ for large $d/(a_0d_c)$ should behave like
$(2/z)\log(d)$. Indeed, as displayed in the figure, the slope of the
curves approaches $4/3$ for large $d/(a_0d_c)$.

%%%%%%%%%%%%%%%%%%%%%%%%%%%%%%%%%%%%%%%%%%%%%%%%%%%%%%%%%%%%%%%%%%%%%%%%%%%%

\section{Conclusions and Further work}
\label{sec:conclusions}

In this work we have elucidated the role of what we have termed
\emph{geodesic degeneracy}, an intrinsic property of the lattice
connectivity, on the scaling behavior of the first-passage percolation
model in discrete planar lattices. We have considered both regular and
disordered lattices, using as examples the square lattice and Delaunay
lattices over uniformly scattered random points.

With regard to the asymptotic scaling behavior, a thorough analysis of
both regular and disordered systems clearly support the conjecture
(not yet proved rigorously) that the FPP model belongs to the KPZ
universality class. Our study covers the fluctuations in the times of
arrival, the geodesic lengths and their lateral deviations, explicitly
demonstrating universal behavior with respect to the type of lattice
and crossing-time distribution functions.

On the other hand, the pre-asymptotic behavior strongly depends on the
geodesic degeneracy of the lattice direction of growth. When the
geodesic in the homogeneous limit is unique or \emph{weakly}
degenerate, a pre-asymptotic trivial scaling corresponding to the
linear growth along that direction is obtained. In this regime, the
geodesic path follows the minimal path of the clean case until arrival
time fluctuations are sufficiently strong to break that geometrical
constraint. At this point the minimal path initiates a strongly
fluctuating behavior that follows the expected KPZ scaling. On the
contrary, if the minimal path is \emph{strongly} degenerate in the
homogeneous case, the geodesic path does not have such topological
constraints and is free to fluctuate from the very beginning. Then KPZ
scaling is rapidly approached. Although some heuristic arguments to
characterize the degree of the geodesic degeneracy were provided, we
are aware that a formal definition remains to be provided.

Interestingly, the crossover from linear growth to KPZ scaling along
weakly degenerate directions seems to be uniquely determined by the
coefficient of variation (CV) of the link-time distribution,
regardless of the ordered or disordered nature of the underlying
lattice. The crossover length and thus the extent of the
pre-asymptotic regime is inversely proportional to the squared
dispersion of the link-times: $d_c\sim\mbox{CV}^{-2}$.

The dependence of the geodesic degeneracy on the lattice direction
leads to anisotropic growth on regular lattices, where an angular
dependence of the degree of geodesic degeneracy occurs. This
anisotropy is reflected in the kinetics and shape of the growing
geodesic balls. Stronger disorder in the link-time distribution
unavoidably entails faster growth than in the homogeneous
case. However, this effect is favored by the degeneracy of the lattice
direction. Again, the coefficient of variation emerges as a key
parameter by determining the velocity of growth along the different
lattice directions and hence the asymptotic shape of the geodesic
ball. For irregular lattices with isotropic disorder we can
accordingly expect isotropic growth.

The preasymptotic regime of FPP ($d\ll d_c$) for a strongly degenerate
direction bears a strong similarity to a directed polymer in a random
medium (DPRM) at zero temperature \cite{Kardar_87,Krug_91,Halpin_95},
if the rightwards and upwards links attached to any site are forced to
be equal. Indeed, the allowed directed polymer configuration coincides
with what we called degenerate paths. Nonetheless, for $d>d_c$, the
geodesic can fold back, and the set of allowed geodesics becomes much
larger than the set of allowed directed polymer
configurations. Interestingly, both of them will be ruled by the KPZ
class.

We should end by stressing that we have dealt only with distributions
with $\mbox{CV}<1$ ($d_c>1$). For distributions with a stronger
dispersion the pre-asymptotic behavior is not observed and different
interesting effects arise which deserve additional investigation and
remained beyond the scope of the present work.

\begin{acknowledgments}
We acknowledge fruitful conversations with E. Korutcheva. This work
has been supported by the Spanish government through MINECO grants
FIS2015-66020-C2-1-P and FIS2015-69167-C2-1-P.
\end{acknowledgments}

%%%%%%%%%%%%%%%%%%%%%%%%%%%%%%%%%%%%%%%%%%%%%%%%%%%%%%%%%%%%%%%%%%%%%%%%%%%%


\begin{thebibliography}{10}

\bibitem{Adler} R. Adler and J. Taylor, {\em Random fields and
  geometry}, Springer (2007).

\bibitem{Itzykson_Drouffe} C. Itzykson and J.-M. Drouffe, {\em
  Statistical Field Theory}, Cambridge University Press (1991).

\bibitem{Booss.book} B. Boo\ss-Bavnbek, G. Esposito and M. Lesch, {\em
  New paths towards quantum gravity}, Springer (2009).

\bibitem{Nelson} D. Nelson, T. Piran and S. Weinberg, {\em
  Statistical Mechanics of Membranes and Surfaces} World Scientific,
  Singapore (2004).

\bibitem{Boal} D. H. Boal, {\em Mechanics of the cell}, Cambridge
  University Press (2012).

\bibitem{Ambjorn_97} J. Ambj{\o}rn, B. Durhuus and T. Jonsson, {\em
  Quantum Geometry: A Statistical Field Theory Approach}, Cambridge
  University Press (1997).

\bibitem{Santalla_15} S.N. Santalla, J. Rodriguez-Laguna, T. LaGatta,
  R. Cuerno, New J. Phys. {\bf 17} 033018 (2015).

\bibitem{Kardar_PRL86} M. Kardar, G. Parisi, Y.-C. Zhang,
  Phys. Rev. Lett. {\bf 56}, 889 (1986).

\bibitem{Barabasi} A.-L. Barab\'asi and H. E. Stanley, {\em Fractal
  Concepts in Surface Growth}, Cambridge University Press (1995).

\bibitem{Krug_97} J. Krug, Adv. Phys. {\bf 46}, 139 (1997).

\bibitem{Krug_10} T. Kriecherbauer, J. Krug, J. Phys. A:
  Math. Theor. {\bf 43}, 403001 (2010).
 
\bibitem{Halpin_15} T. Halpin-Healy, K. Takeuchi, J. Stat. Phys. {\bf
  160}, 794 (2015).
  
\bibitem{Takeuchi_11} K. A. Takeuchi, M. Sano, T. Sasamoto,
  H. Spohn, Sci. Rep. {\bf 1}, 34 (2011).

\bibitem{Praehofer_02} M. Pr\"ahofer, H. Spohn, J. Stat. Phys. {\bf
  108}, 1071 (2002).

\bibitem{Corwin_13} I. Corwin, J. Quastel, D. Ramenik,
  Comm. Math. Phys. {\bf 317}, 347 (2013).

\bibitem{Santalla_17} S.N. Santalla, J. Rodriguez-Laguna, A. Celi,
  R. Cuerno, J. Stat. Mech. 023201 (2017).

\bibitem{Hammersley_65} J. M. Hammersley and D. J. A. Welsh,
  ``First-passage percolation, subadditive processes, stochastic
  networks and generalized renewal theory'', in {\em Bernoulli, Bayes,
    Laplace anniversary volume}, J. Neyman and L. M. LeCam eds.,
  Springer-Verlag (1965), p. 61.

\bibitem{Howard_04} C. D. Howard, ``Models of first passage
  percolation'', in {\em Probability on discrete structures},
  H. Kesten ed., Springer (2004), p. 125--173.

\bibitem{Auffinger_17} A. Auffinger, M. Damron, J. Hanson, {\em 50
  years of FPP}, University Lecture Series 68, American Mathematical
  Society (2017).

\bibitem{Smythe_78} R.T. Smythe, J.C. Wierman, {\em First-passage
  percolation on the square lattice}, Springer Verlag (1978).
  
\bibitem{Ritzenberg_84} A.L. Ritzenberg, R.J. Cohen, Phys. Rev. B {\bf
  30}, 4038 (1984).
  
\bibitem{Kerstein_86} A.R. Kerstein, B.F. Edwards, Phys. Rev. B {\bf
  33} 3353 (1986).
  
\bibitem{Yao_13} C.-L. Yao, Stat. \&\ Prob. Lett. {\bf 83}, 797 (2013).
  
\bibitem{Yao_18} C.-L. Yao, Stoch. Proc. Appl. {\bf 128}, 445 (2018).

\bibitem{Fill_93} J.A. Fill, R. Pemantle, Ann. Appl. Probab. {\bf 3},
  593 (1993)

\bibitem{Bhamidi_10} S. Bhamidi, R. van der Hofstad, G. Hooghiemstra,
  Ann. Appl. Probab. {\bf 20}, 1907 (2010).

\bibitem{Chatterjee_13} S. Chatterjee, Ann. Math. {\bf 177}, 663
  (2013).

\bibitem{Kardar_87} M. Kardar, Y.-C. Zhang, Phys. Rev. Lett. {\bf 58},
  2087 (1987).

\bibitem{Krug_91} J. Krug, H. Spohn, in {\em Solids far from
  equilibrium}, Ed. C. Godr\`eche, Cambridge University Press (1991).
  
\bibitem{Halpin_95} T. Halpin-Healy, Y.-C. Zhang, Phys. Rep. {\bf
  254}, 215 (1995).
  
\bibitem{Abraham_95} D.B. Abraham, L. Fontes, C.W. Newman,
  M.S.T. Piza, Phys. Rev. E {\bf 52}, R1257 (1995).

\bibitem{Beyme_14} S. Beyme, C. Leung, Ad Hoc Netw. {\bf 17}, 60
  (2014).
  
\bibitem{Kordzhakia_05} G. Kordzakhia, S.P. Lalley,
  Stoch. Proc. App. {\bf 115}, 781 (2005).

\bibitem{Bundschuh_00} R. Bundschuh, T. Hwa, Discr. Appl. Math. {\bf
  104}, 113 (2000).
  
\bibitem{Cormen} T.H. Cormen, C.E. Leiserson, R.L. Rivest, C. Stein,
  {\em Introduction to algorithms}, The MIT Press (1990).

\bibitem{Richardson_73} D. Richardson, Proc. Cambridge Phil. Soc. {\bf
  74}, 515 (1973).

\bibitem{Cox-Durrett_81} J. T. Cox, R. Durrett, Ann. Probab. {\bf 9},
  583 (1981).

\bibitem{Kesten_86} H. Kesten, ``Aspects of First-Passage
  Percolation'' in Lect. Notes in Mathematics {\bf 1180},
  125-264. Springer Verlag (1986).

\bibitem{Bowyer_81} A. Bowyer, Comput. J. {\bf 24}, 162 (1981).

\bibitem{Watson_81} D.F. Watson, Comput. J. {\bf 24}, 167 (1981).

\bibitem{Rebay_93} S. Rebay, J. Comp. Phys. {\bf 106}, 127 (1993).
  
\bibitem{Vasicek.76} O. Vasicek, J. Roy. Statist. Soc. Ser. B {\bf
  38}, 54 (1976).

\bibitem{Desurvire_09} E. Desurvire, {\em Classical and quantum information
  theory}, Cambridge University Press (2009).

  
\bibitem{note_entropy} We should emphasize that we do not measure the
  {\em differential entropy}, because it is not invariant under
  reparametrizations. Instead, we obtain the relative entropy, also
  called Kullback-Leibler divergence \cite{Desurvire_09}, with respect
  to a homogeneous probability distribution in a fixed interval. The
  same reference distribution should be used for all measurements in
  order to be compared.


\end{thebibliography}
\end{document}